\documentclass[journal,onecolumn]{IEEEtran}
\usepackage{amsmath, epsfig, cite}
\usepackage{amssymb}
\usepackage{amsfonts}
\usepackage{latexsym}
\usepackage{longtable}
\usepackage{tabularx}
\usepackage{color}
\usepackage{multirow}

\newtheorem{theorem}{Theorem}

\title{Private Information Retrieval from MDS Coded Databases with Colluding Servers under Several Variant Models}

\author{Yiwei Zhang\thanks{The research of Y. Zhang was supported by the Post-Doctoral Science Foundation of China under Grant No. 2017M610942 and Beijing Postdoctoral Research Foundation.} and Gennian Ge
\thanks{The research of G. Ge was supported by the National Natural Science Foundation of China under Grant Nos.11431003 and 61571310, Beijing Scholars
Program, Beijing Hundreds of Leading Talents Training Project of Science and Technology, and Beijing Municipal
Natural Science Foundation.}
\thanks{Y. Zhang is with the School of Mathematical Sciences, Capital Normal University, Beijing 100048, China (email: rexzyw@163.com).}
\thanks{G. Ge is with the School of Mathematical Sciences, Capital Normal University, Beijing 100048, China (e-mail: gnge@zju.edu.cn).}
}

\begin{document}

\date{}\maketitle

\begin{abstract}
Private information retrieval (PIR) gets renewed attentions due to its information-theoretic reformulation and its application in distributed storage system (DSS). The general PIR model considers a coded database containing $N$ servers storing $M$ files. Each file is stored independently via the same arbitrary $(N,K)$-MDS code. A user wants to retrieve a specific file from the database privately against an arbitrary set of $T$ colluding servers. A key problem is to analyze the PIR capacity, defined as the maximal number of bits privately retrieved per one downloaded bit.  Several extensions for the general model appear by bringing in various additional constraints. In this paper, we propose a general PIR scheme for several variant PIR models including: PIR with robust servers, PIR with Byzantine servers, the multi-file PIR model and PIR with arbitrary collusion patterns.
\end{abstract}


\section{Introduction} \label{SecIntro}

The concept of private information retrieval is first introduced by Chor et al. \cite{Chor1,Chor2}. In a system consisting of several servers each storing the same $n$-bit database, a classic PIR scheme will allow a user to retrieve a single bit of the database without revealing the identity of the desired bit to any single server. PIR has received a lot of attention from the computer science community and cryptography community. The main research objective is to minimize the total communication cost, including both the upload cost (the queries sent to the servers) and the download cost (the feedbacks from the servers). A series of works have managed to reduce the communication cost, see for example, \cite{Beimel,Efremenko,Dvir,Yekhanin,YekhaninSurvey} and the references therein.


Recently, the PIR problem is reformulated from an information-theoretic point of view. Instead of retrieving a single bit from an $n$-bit database, the information-theoretic PIR problem considers retrieving a single file from an $n$-file database, where the length of each file could be arbitrarily large. In this sense, the objective turns into minimizing the total communication cost per unit of retrieved data. It is further noticed in \cite{Chan} that in this information-theoretic model the upload cost could be neglected with respect to the download cost. Meanwhile, with the development of distributed storage system (DSS), it is natural to consider designing PIR schemes for coded databases instead of the replicated-based databases  \cite{Augot,Shah,Chan,Blackburn}.

The general model for the information-theoretic, DSS-oriented PIR problem is as follows. Assume we have a coded database containing $N$ servers storing $M$ files. Each file is stored independently via the same arbitrary $(N,K)$-MDS code. A user wants to retrieve a specific file from the database. Any $T$ out of the $N$ servers may collude in attacking on the privacy of the user. We call the PIR scheme for such a system as an $(N,K,T;M)$-scheme. PIR rate is defined as the number of bits of messages that could be privately retrieved per one bit of downloaded message. The supremum of all achievable rates is called the PIR capacity, denoted by $C=C(N,K,T;M)$. Starting from the pioneering work \cite{Sun} by Sun and Jafar, a series of works have attempted to analyze the fundamental limits on the PIR capacity. The exact value of the PIR capacity for the degenerating cases, i.e., either $K=1$ or $T=1$, has been completely solved \cite{BanawanCoded,SunColluded}. For MDS coded databases with colluding servers ($K\ge2$ and $T\ge2$), determining the capacity still remains an open problem and several PIR schemes for the general model can be found in \cite{Freij-Hollanti,SunNew,TajeddineConfversion,TajeddineFullversion,Zhangnew}.

Several extensions for the general PIR model gradually appear by bringing in various additional constraints. For example, the symmetric PIR problem is considered in \cite{SunSymmetric}, \cite{Wang} and \cite{Wang1}, where a further constraint is that the user should learn nothing about any undesired file. \cite{SunMultiround} discusses multi-round PIR schemes, where the queries and feedbacks are made in several rounds and the user may adjust his queries according to the feedbacks from previous rounds. \cite{Kumar} considers the model replacing the underlying MDS storage code by some non-MDS code. Recent months have witnessed a new trend of researches on PIR with side information \cite{Tandon,Kadhe,Chen,Wei,Wei1}. It is expected that more variant models of PIR will arise according to different practical scenarios.

In this paper, we will mainly consider the following four PIR variant models:

$\bullet$ 1. \emph{PIR with robust servers}. In this model, some $S$ out of the $N$ servers are robust servers and they fail to respond to the user's queries. Thus this can be thought of as an erasure-correcting PIR model. PIR with robust servers has been studied by Sun and Jafar in \cite{SunColluded} for the degenerating case $K=1$ and then by Tajeddine and Rouayheb in \cite{TajeddineRobust} for the degenerating case $T=1$.

$\bullet$ 2. \emph{PIR with Byzantine servers}. In this model, among the $N$ servers we have $B$ Byzantine servers, which respond with incorrect answers due to their outdated contents or even adversary motivations. Thus this can be thought of as an error-correcting PIR model. For the degenerating case $K=1$, PIR with Byzantine servers has been studied by Banawan and Ulukus in \cite{BanawanByzantine}.

$\bullet$ 3. \emph{Multi-file PIR}. This model assumes that a user wants to retrieve several files simultaneously. Multi-file PIR is first considered by Banawan and Ulukus \cite{BanawanMultimessage} for the case $K=T=1$. They show that in order to retrieve $P\ge2$ files, one can do better than the trivial approach of executing $P$ independent (single-file) PIR schemes.

$\bullet$ 4. \emph{PIR with arbitrary collusion patterns}. Instead of the original ``any-$T$-out-of-the-$N$-servers-may-collude" setting, in this model we are given a collusion pattern, which is a family of subsets of the servers closed under inclusion. A subset in the family indicates that this set of servers may collude in attacking on the privacy. This model is first proposed in \cite{TajeddineNew}, which mainly analyzes the effectiveness of the general PIR scheme proposed in \cite{Freij-Hollanti} in the new scenario.

In this paper, we will consider the first three models above in MDS-coded databases with colluding servers, i.e., we study these models for non-degenerated parameters $K\ge2$ and $T\ge2$. The main approach is to adapt the general PIR scheme we presented in \cite{Zhangnew} according to the additional constraints in different models. Also we will investigate the effectiveness of our previous strategy for the model with arbitrary collusion patterns. A summary of the results for each independent model are listed as follows.

$\bullet$ 1. \emph{PIR with robust servers}. When $T+K\le N$ and ${{N-S}\choose K}>{{N}\choose K}-{{N-T}\choose K}$, we have an $(N,K,T;M)$-PIR scheme with $S$ robust servers whose rate is $\frac{{{N-S-1}\choose{K-1}}}{{{N-1}\choose{K-1}}}(1+R+R^2+\cdots+R^{M-1})^{-1}$, where $R=\frac{{N\choose K}-{{N-T}\choose K}}{{{N-S}\choose K}}$. This result contains the degenerating case $K=1$ presented in \cite{SunColluded} as a special case.

$\bullet$ 2. \emph{PIR with Byzantine servers}. When $T+K\le N$ and $2{{N-B}\choose{K}}-{{N}\choose{K}}>{{N}\choose K}-{{N-T}\choose K}$, we have an $(N,K,T;M)$-PIR scheme with $B$ robust servers whose rate is $\frac{2{{N-B}\choose{K}}-{{N}\choose{K}}}{{N\choose K}}(1+R+R^2+\cdots+R^{M-1})^{-1}$, where $R=\frac{{N\choose K}-{{N-T}\choose K}}{2{{N-B}\choose{K}}-{{N}\choose{K}}}$. This result contains the degenerating case $K=1$ presented in \cite{BanawanByzantine} as a special case.

$\bullet$ 3. \emph{Multi-file PIR}. In this part, we first present an upper bound of the multi-file PIR capacity for the degenerating cases, i.e., one of the parameters $K$ and $T$ equals one. We then present a multi-file PIR scheme for the case $\frac{P}{M}\ge\frac{1}{2}$, which works for general parameters $K$ and $T$ and is of optimal PIR rate for the degenerating cases.

$\bullet$ 4. \emph{PIR with arbitrary patterns}. We adapt our PIR scheme in this model and its PIR rate depends on an optimization problem related to combinatorial design theory.

The rest of the paper is organized as follows. In Section \ref{SecModel}, we introduce the general model of the PIR problem alongside with the different settings with respect to each variant model. In Section \ref{Prototype} we briefly recall the general PIR scheme we presented in \cite{Zhangnew}, which is the prototype of all the upcoming schemes for the variant models. Sections \ref{SecRobust}, \ref{SecByzantine}, \ref{SecMulti} and \ref{SecArbitrary} present the main results of the paper, analyzing the four models respectively. Finally Section \ref{SecConclusion} concludes the paper.

\section{Problem Statement}\label{SecModel}

In this section we introduce the general model of the PIR problem alongside with the different settings with respect to each variant model. Basically the statement follows the same way as shown in \cite{BanawanMultimessage,BanawanCoded,Sun,SunColluded,SunNew,Zhangnew}.

The general model considers a distributed storage system consisting of $N$ servers. The system stores $M$ files, denoted as $W^{[1]},W^{[2]},\dots,W^{[M]}\in\mathbb{F}_q^{L\times K}$, i.e., each file is of length $LK$ and represented in a matrix of size $L\times K$. The files are independent and identically distributed with
\begin{align}
H(W^{[i]})&=LK,~~~i\in\{1,2,\dots,M\},\\
H(W^{[1]},W^{[2]},\dots,W^{[M]})&=MLK.
\end{align}

Denote the $j$th row of the file $W^{[i]}$ as $\mathbf{w}_j^{[i]}\in\mathbb{F}_q^{K}$, $1\le j \le L$. Each file is stored in the system via the same given $(N,K)$-MDS code. The generator matrix $\mathbf{G}\in \mathbb{F}_q^{K\times N}$ of the MDS code is denoted as
\begin{equation}
\mathbf{G}=\Big[ \mathbf{g_1}~~\mathbf{g_2}~~\cdots~~\mathbf{g_N} \Big]_{K\times N},
\end{equation}
and the MDS property means that any $K$ columns of $\mathbf{G}$ are linearly independent. For each $\mathbf{w}_j^{[i]}$, the $n$th server stores the coded bit $\mathbf{w}_j^{[i]}\mathbf{g_n}$. Thus, the whole contents $\mathbf{y}_n\in\mathbb{F}_q^{ML}$ stored on the $n$th server are the concatenated projections of all the files $\{W^{[1]},W^{[2]},\dots,W^{[M]}\}$ on the encoding vector $\mathbf{g_n}$, i.e.,
\begin{align}
  \mathbf{y}_n&=\left(
                  \begin{array}{c}
                    W^{[1]} \\
                    \vdots \\
                    W^{[M]} \\
                  \end{array}
                \right)\mathbf{g_n}\\
  &=\Big[ \mathbf{w}_1^{[1]}\mathbf{g_n}~~\cdots~~\mathbf{w}_L^{[1]}\mathbf{g_n}~~\mathbf{w}_1^{[2]}\mathbf{g_n}~~\cdots~~\mathbf{w}_L^{[2]}\mathbf{g_n}~~\cdots~~\mathbf{w}_1^{[M]}\mathbf{g_n}~~\cdots~~\mathbf{w}_L^{[M]}\mathbf{g_n} \Big]^T.
\end{align}

Now assume that a user wants to retrieve an arbitrary file $W^{[i]}$. The retrieval is done by sending some queries to the servers and then analyzing the feedbacks. Let $\mathcal{F}$ denote a random variable generated by the user and unknown to any server. $\mathcal{F}$ represents the randomness of the user's strategy to generate the queries. Let $\mathcal{G}$ denote a random variable generated by the servers and also known to the user. $\mathcal{G}$ represents the randomness of the server's strategy to produce the feedbacks\footnote{The role of $\mathcal{G}$ is first proposed by Sun and Jafar in \cite{SunNew}. Such a strategy allows the server to perform some coding procedures before responding and thus may reduce the download cost. For more details please refer to \cite{SunNew}. In almost all the other works on PIR schemes, including the current draft, we follow a question-and-answer format. That is, for any query vector the server responds the corresponding projection of his contents onto the query. For this kind of schemes $\mathcal{G}$ is a deterministic strategy and thus can be neglected.}. Both $\mathcal{F}$ and $\mathcal{G}$ are generated independently of the files and the index $i$ of the desired file, i.e.,
\begin{equation}
  H(\mathcal{F},\mathcal{G},i,W^{[1]},W^{[2]},\dots,W^{[M]})=H(\mathcal{F})+H(\mathcal{G})+H(i)+H(W^{[1]})+\cdots+H(W^{[M]}).
\end{equation}

Using his strategy $\mathcal{F}$, the user generates a set of queries $Q^{[i]}_n$ to the $n$th server, $1\le n \le N$. The queries are independent of the files, i.e.,
\begin{equation}
  \text{[Query]}~~I(Q^{[i]}_1,Q^{[i]}_2,\dots,Q^{[i]}_N;W^{[1]},W^{[2]},\dots,W^{[M]})=0.
\end{equation}

Upon receiving the query, the $n$-th server responds a feedback $A^{[i]}_n$, which is a deterministic function of the query $Q^{[i]}_n$, the strategy $\mathcal{G}$ and the data $\mathbf{y}_n$ (and therefore a deterministic function of the query $Q^{[i]}_n$, the strategy $\mathcal{G}$ and the files $\{W^{[1]},W^{[2]},\dots,W^{[M]}\}$), i.e.,
\begin{equation} \label{FeedbackConstraint}
  \text{[Feedback]}~~H(A^{[i]}_n|Q^{[i]}_n,\mathcal{G},\mathbf{y}_n)=H(A^{[i]}_n|Q^{[i]}_n,\mathcal{G},W^{[1]},W^{[2]},\dots,W^{[M]})=0.
\end{equation}

The user retrieves his desired file $W^{[i]}$ based on all the queries and feedbacks, plus the knowledge of the strategies $\mathcal{F}$ and $\mathcal{G}$, i.e.,
\begin{equation} \label{CorrectnessConstraint}
  \text{[Correctness]}~~H(W^{[i]}|Q^{[i]}_1,Q^{[i]}_2,\dots,Q^{[i]}_N,A^{[i]}_1,A^{[i]}_2,\dots,A^{[i]}_N,\mathcal{F},\mathcal{G})=0.
\end{equation}

For any subset $\mathcal{T}$ of the servers, $|\mathcal{T}|=T$, let $Q^{[i]}_\mathcal{T}$ represent $\{Q^{[i]}_n,n\in\mathcal{T}\}$. Similarly we have the notation $A^{[i]}_\mathcal{T}$. The privacy constraint requires that this set of servers learns nothing about the identity of the retrieved file, i.e.,
\begin{equation} \label{PrivacyConstraint}
  \text{[Privacy]}~~I(i;Q^{[i]}_\mathcal{T})=I(i;A^{[i]}_\mathcal{T})=0, ~\forall\mathcal{T}\subseteq\{1,2,\dots,N\},~|\mathcal{T}|=T.
\end{equation}

The rate for the PIR scheme is defined as the ratio of the size of the retrieved file to the total download cost, i.e.,
\begin{equation}
  \frac{H(W^{[i]})}{\sum_{n=1}^N H(A^{[i]}_n)},
\end{equation}
and the PIR capacity $C$ is the supremum of all achievable rates.

For each variant model, the statement above requires slight modifications.

\subsection{Problem Statement for PIR with robust servers}

In this model, some $S$ out of the $N$ servers may fail to respond. Under this assumption, in order to successfully retrieve a file, the correctness constraint (\ref{CorrectnessConstraint}) should be modified as
\begin{equation}
  \text{[Correctness with robust servers]}~~H(W^{[i]}|Q^{[i]}_1,Q^{[i]}_2,\dots,Q^{[i]}_N,A^{[i]}_{j_1},A^{[i]}_{j_2},\dots,A^{[i]}_{j_{N-S}},\mathcal{F},\mathcal{G})=0,
\end{equation}
where $\mathcal{J}=\{j_1,j_2,\dots,j_{N-S}\}$ is an arbitrary subset of $[N]$. The complement set $\overline{\mathcal{J}}$ represents those $S$ robust servers that fail to respond. Note that a robust server can also collude in attacking on the privacy.

\subsection{Problem Statement for PIR with Byzantine servers}

In this model, there exists a set $\mathcal{B}$ of Byzantine servers unknown to the user. $|\mathcal{B}|=B$. These servers will respond incorrect answers due to outdated contents or even adversary motivations. That is, for a Byzantine server, the feedback (\ref{FeedbackConstraint}) should be modified as
\begin{equation}
  \text{[Feedback from Byzantine servers]}~~H(A^{[i]}_n|Q^{[i]}_n,\mathcal{G},\mathbf{y}_n)=H(A^{[i]}_n|Q^{[i]}_n,\mathcal{G},W^{[1]},W^{[2]},\dots,W^{[M]})\ge0,~n\in\mathcal{B}.
\end{equation}

Note that as remarked in \cite{BanawanByzantine}, we should differentiate the actions of $T$ servers colluding in attacking on the privacy and $B$ servers coordinating in introducing errors. No specific relation between these two actions are assumed.

\subsection{Problem Statement for multi-file PIR}

In this model, a user wants to retrieve an arbitrary set of $P\ge 2$ files simultaneously and the set of indices of these files is $\mathcal{P}=\{i_1,i_2,\dots,i_P\}\subset\{1,2,\dots,M\}$. Denote $W^{[\mathcal{P}]}=\{W^{[i]},i\in\mathcal{P}\}$. To retrieve $W^{[\mathcal{P}]}$, the queries and feedbacks are then denoted as $Q^{[\mathcal{P}]}_n$ and $A^{[\mathcal{P}]}_n$, $n\in[N]$. The general statement should be adjusted to a multi-file version as follows.
\begin{equation}
  \text{[Multi-file Query]}~~I(Q^{[\mathcal{P}]}_1,Q^{[\mathcal{P}]}_2,\dots,Q^{[\mathcal{P}]}_N;W^{[1]},W^{[2]},\dots,W^{[M]})=0.
\end{equation}
\begin{equation}
  \text{[Multi-file Feedback]}~~H(A^{[\mathcal{P}]}_n|Q^{[\mathcal{P}]}_n,\mathcal{G},\mathbf{y}_n)=H(A^{[\mathcal{P}]}_n|Q^{[\mathcal{P}]}_n,\mathcal{G},W^{[1]},W^{[2]},\dots,W^{[M]})=0.
\end{equation}
\begin{equation}
  \text{[Multi-file Correctness]}~~H(W^{[\mathcal{P}]}|Q^{[\mathcal{P}]}_1,Q^{[\mathcal{P}]}_2,\dots,Q^{[\mathcal{P}]}_N,A^{[\mathcal{P}]}_1,A^{[\mathcal{P}]}_2,\dots,A^{[\mathcal{P}]}_N,\mathcal{F},\mathcal{G})=0.
\end{equation}
\begin{equation}
  \text{[Multi-file Privacy]}~~I(\mathcal{P};Q^{[\mathcal{P}]}_\mathcal{T})=I(\mathcal{P};A^{[\mathcal{P}]}_\mathcal{T})=0, ~\forall\mathcal{T}\subseteq\{1,2,\dots,N\},~|\mathcal{T}|=T.
\end{equation}
Correspondingly, the multi-file version PIR rate is defined as
\begin{equation}
  \frac{\sum_{i\in\mathcal{P}}H(W^{[i]})}{\sum_{n=1}^N H(A^{[\mathcal{P}]}_n)},
\end{equation}
and the multi-file PIR capacity $C^P$ is the supremum of all achievable rates.

\subsection{Problem Statement for PIR with arbitrary collusion patterns}

In this model, we have a collusion pattern $\mathfrak{T}$, which is a family of subsets of $[N]$ closed under inclusion. A subset $\mathcal{T}\in\mathfrak{T}$ indicates that this set of servers may collude in attacking on the identity of the desired file. So the privacy constraint (\ref{PrivacyConstraint}) should be modified as
\begin{equation}
  \text{[Privacy against collusion pattern $\mathfrak{T}$]}~~I(i;Q^{[i]}_\mathcal{T})=I(i;A^{[i]}_\mathcal{T})=0, ~\forall\mathcal{T}\in\mathfrak{T}.
\end{equation}

\section{The prototype: a PIR scheme for the general model} \label{Prototype}

In this section we recall the general PIR scheme we proposed in \cite{Zhangnew}. It is the prototype of the upcoming PIR schemes for all the variant models. The scheme suits the general model with $T+K\le N$ and achieves the best rate up to now for a certain range of parameters. For a detailed comparison of several PIR schemes for the general PIR model, please refer to \cite[Section IV]{Zhangnew}.

\begin{theorem}\text{\cite[Theorem 1]{Zhangnew}} \label{OriginalRate}
  When $T+K\le N$, there exists an $(N,K,T;M)$-PIR scheme with rate
  $$(1+R+R^2+\cdots+R^{M-1})^{-1}, \text{ where }R=1-\frac{{{N-T}\choose K}}{{N\choose K}}.$$
\end{theorem}

\subsection{The general scheme}

We repeat the notations again as shown in the problem setting and these notations will be used throughout the draft. Denote the files as $W^{[1]},W^{[2]},\dots,W^{[M]}$ and without loss of generality we let $W^{[1]}$ be the desired file. Each file is of length $LK$ and represented in a matrix of size $L\times K$ over $\mathbb{F}_q$, where $L$ is a constant to be determined later. $\mathbb{F}_q$ is a sufficiently large finite field that allows the existence of several MDS codes used in the scheme  (in Step 2). Each file, say $W^{[m]}$, consists of $L$ rows, $\mathbf{w}^{[m]}_1,\mathbf{w}^{[m]}_2,\dots,\mathbf{w}^{[m]}_L$. For each file $W^{[m]}$ we will build a list of ``atoms". Each {\it atom} is a linear combination of $\{\mathbf{w}^{[m]}_1,\mathbf{w}^{[m]}_2,\dots,\mathbf{w}^{[m]}_L\}$, represented in the form of $\mathbf{s}W^{[m]}$, where $\mathbf{s}$ is a vector in $\mathbb{F}_q^{L}$. A query to a server can be written as a linear combination of atoms towards different files and therefore a linear combination of $\{\mathbf{w}^{[m]}_l:1\le m \le M,1\le l \le L\}$. Upon receiving a query, the server responds the corresponding coded bit based on its storage $\{\mathbf{w}^{[m]}_l\mathbf{g}_n:1\le m \le M,1\le l \le L\}$.

Constructing the PIR scheme contains the following steps.

$\bullet$ Step 1: Let $\alpha$ and $\beta$ be the smallest positive integers satisfying
\begin{equation}
\alpha{{N}\choose{K}}=(\alpha+\beta)\Bigg({{N}\choose{K}}-{{N-T}\choose{K}}\Bigg).
\end{equation}
Assume the existence of an $\big((\alpha+\beta){{N}\choose{K}},\alpha{{N}\choose{K}}\big)$-MDS code. The transpose of its generator matrix is denoted as $\text{MDS}_{(\alpha+\beta){{N}\choose{K}}\times\alpha{{N}\choose{K}}}$.

$\bullet$ Step 2: Let $L={{N}\choose{K}}(\alpha+\beta)^{M-1}$. Independently choose $M$ random matrices, $S_1,\dots,S_M$, uniformly from all the $L\times L$ full rank matrices over $\mathbb{F}_q$.

$\bullet$ Step 3: Build an {\it assisting array} of size ${{N-1}\choose{K-1}}\times N$ consisting of ${{N}\choose{K}}$ symbols. Each symbol appears $K$ times and every $K$ columns share a common symbol.

$\bullet$ Step 4: [\emph{Building basic blocks of the query structure}] The query structure is divided into several blocks, where each block is labelled by a nonempty subset of files $\mathcal{D}\subseteq\{W^{[1]},W^{[2]},\dots,W^{[M]}\}$. In a block labelled by $\mathcal{D}$, every query is a mixture of $|\mathcal{D}|$ atoms related to the files in $\mathcal{D}$. We set each block in an ``isomorphic" form with the assisting array, i.e., every $K$ servers share a common query. We further call a block labelled by $\mathcal{D}$ a {\it $d$-block} if $|\mathcal{D}|=d$. For any $\mathcal{D}\subseteq\{W^{[1]},W^{[2]},\dots,W^{[M]}\}$, $|\mathcal{D}|=d$, we require that the number of $d$-blocks labelled by $\mathcal{D}$ is $\alpha^{M-d}\beta^{d-1}$.

$\bullet$ Step 5: \emph{[Constructing the atoms for the desired file in each block]} The atoms for the desired file $W^{[1]}$ are just built by $S_1W^{[1]}$ and then distributed to different blocks.

$\bullet$ Step 6: \emph{[Dividing the query structure into groups]} Now let $\mathcal{D}$ denote a nonempty set of files not containing the desired file $W^{[1]}$. There are totally $\alpha^{M-|\mathcal{D}|}\beta^{|\mathcal{D}|-1}$ blocks labelled by $\mathcal{D}$ and $\alpha^{M-|\mathcal{D}|-1}\beta^{|\mathcal{D}|}$ blocks labelled by $\mathcal{D}\bigcup\{W^{[1]}\}$. Partition these blocks into $\alpha^{M-|\mathcal{D}|-1}\beta^{|\mathcal{D}|-1}$ groups, denoted by $\Gamma^{\mathcal{D}}_{\lambda}$, $1\le\lambda\le \alpha^{M-|\mathcal{D}|-1}\beta^{|\mathcal{D}|-1}$, where each group consists of $\alpha$ blocks labelled by $\mathcal{D}$ and $\beta$ blocks labelled by $\mathcal{D}\bigcup\{W^{[1]}\}$. Repeat this process for every nonempty set $\mathcal{D}$ not containing $W^{[1]}$ and thus the whole query structure is partitioned into groups.

$\bullet$ Step 7: \emph{[Constructing the atoms for undesired files in each group]} For each group constructed with respect to $\mathcal{D}$ and any file $W^{[m]}\in\mathcal{D}$, $m\neq1$, the number of atoms towards $W^{[m]}$ within the blocks in this group is $(\alpha+\beta){{N}\choose{K}}$, among which $\alpha{{N}\choose{K}}$ atoms appear in blocks labelled by $\mathcal{D}$ and the other $\beta{{N}\choose{K}}$ atoms appear as interferences in blocks labelled by $\mathcal{D}\bigcup\{W^{[1]}\}$. Let these $(\alpha+\beta){{N}\choose{K}}$ atoms for $W^{[m]}$ be built by $\text{MDS}_{(\alpha+\beta){{N}\choose{K}}\times\alpha{{N}\choose{K}}}S'_mW^{[m]}$, where $S'_m$ denotes some $\alpha{{N}\choose{K}}$ rows in the matrix $S_m$.

Repeat this process for all the groups. Note that we shall come across the same file $W^{[M]}$ several times. We have ${{M-2}\choose{|\mathcal{D}|-1}}$ choices for a subset $\mathcal{D}$ of size $|\mathcal{D}|$, containing $W^{[M]}$ but without $W^{[1]}$. With respect to any such $\mathcal{D}$ we have $\alpha^{M-|\mathcal{D}|-1}\beta^{|\mathcal{D}|-1}$ groups. So the exact number of times we come across $W^{[M]}$ can be computed as
\begin{equation}
  \sum_{|\mathcal{D}|=1}^{M-1} \alpha^{M-|\mathcal{D}|-1}\beta^{|\mathcal{D}|-1}{{M-2}\choose{|\mathcal{D}|-1}}=(\alpha+\beta)^{M-2}.
\end{equation}

Every time we come across the same file $W^{[M]}$, we have to ensure that the rows we select from $S_m$ are non-intersecting. This can be satisfied since $(\alpha+\beta)^{M-2}\alpha{{N}\choose{K}}\le L$.

\subsection{An example: $N=4$, $K=2$, $T=2$ and $M=3$}

We illustrate the scheme via an example: $N=4$, $K=2$, $T=2$ and $M=3$. The parameters throughout the construction can be calculated as $\alpha=5$, $\beta=1$ and $L=216$. The assisting array could be of the form
$$\Bigg\{\begin{array}{cccc}
  1 & 1 & 2 & 2 \\
  3 & 4 & 3 & 4 \\
  5 & 6 & 6 & 5
\end{array}\Bigg\}.$$ Now we have three files $U$, $V$ and $W$. Let each file be of length $432$ and represented in a matrix of size $216\times2$, i.e.,
\begin{equation}
  U=\Bigg(
                  \begin{array}{c}
                    \mathbf{u}_1 \\
                    \vdots \\
                    \mathbf{u}_{216} \\
                  \end{array}
                \Bigg)
  \hspace{3em}
  V=\Bigg(
                  \begin{array}{c}
                    \mathbf{v}_1 \\
                    \vdots \\
                    \mathbf{v}_{216} \\
                  \end{array}
                \Bigg)
  \hspace{3em}
  W=\Bigg(
                  \begin{array}{c}
                    \mathbf{w}_1 \\
                    \vdots \\
                    \mathbf{w}_{216} \\
                  \end{array}
                \Bigg)
\end{equation}
where $\mathbf{u}_i,\mathbf{v}_i,\mathbf{w}_i\in\mathbb{F}_q^{2}$, $1\le i \le 216$. Here $\mathbb{F}_q$ is a sufficiently large finite field. Let $U$ be the desired file. We shall construct three lists of atoms $a_{[1:216]}$, $b_{[1:216]}$ and $c_{[1:216]}$, corresponding to each of the files accordingly. Following Step 4, the query structure is divided into several blocks as follows. In each block we set the queries in an ``isomorphic" form with the assisting array.

$$\Lambda^{UVW}:\Bigg\{\begin{array}{cccc}
  \text{Server I} & \text{Server II} & \text{Server III} & \text{Server IV}\\\hline
  a_{1}+b_{1}+c_{1} & a_{1}+b_{1}+c_{1} & a_{2}+b_{2}+c_{2} & a_{2}+b_{2}+c_{2} \\
  a_{3}+b_{3}+c_{3} & a_{4}+b_{4}+c_{4} & a_{3}+b_{3}+c_{3} & a_{4}+b_{4}+c_{4} \\
  a_{5}+b_{5}+c_{5} & a_{6}+b_{6}+c_{6} & a_{6}+b_{6}+c_{6} & a_{5}+b_{5}+c_{5} \\
\end{array}\Bigg\}\text{,}$$

$$\Lambda^{VW}_{\lambda}:\Bigg\{\begin{array}{cccc}
  \text{Server I} & \text{Server II} & \text{Server III} & \text{Server IV}\\\hline
  b_{6\lambda+1}+c_{6\lambda+1} & b_{6\lambda+1}+c_{6\lambda+1} & b_{6\lambda+2}+c_{6\lambda+2} & b_{6\lambda+2}+c_{6\lambda+2} \\
  b_{6\lambda+3}+c_{6\lambda+3} & b_{6\lambda+4}+c_{6\lambda+4} & b_{6\lambda+3}+c_{6\lambda+3} & b_{6\lambda+4}+c_{6\lambda+4} \\
  b_{6\lambda+5}+c_{6\lambda+5} & b_{6\lambda+6}+c_{6\lambda+6} & b_{6\lambda+6}+c_{6\lambda+6} & b_{6\lambda+5}+c_{6\lambda+5} \\
\end{array}\Bigg\} \text{ for $\lambda\in\{1,2,3,4,5\}$,}$$

$$\Lambda^{UV}_{\lambda}:\Bigg\{\begin{array}{cccc}
  \text{Server I} & \text{Server II} & \text{Server III} & \text{Server IV}\\\hline
  a_{6\lambda+1}+b_{6\lambda+31} & a_{6\lambda+1}+b_{6\lambda+31} & a_{6\lambda+2}+b_{6\lambda+32} & a_{6\lambda+2}+b_{6\lambda+32} \\
  a_{6\lambda+3}+b_{6\lambda+33} & a_{6\lambda+4}+b_{6\lambda+34} & a_{6\lambda+3}+b_{6\lambda+33} & a_{6\lambda+4}+b_{6\lambda+34} \\
  a_{6\lambda+5}+b_{6\lambda+35} & a_{6\lambda+6}+b_{6\lambda+36} & a_{6\lambda+6}+b_{6\lambda+36} & a_{6\lambda+5}+b_{6\lambda+35} \\
\end{array}\Bigg\}\text{ for $\lambda\in\{1,2,3,4,5\}$,}$$

$$\Lambda^{UW}_{\lambda}:\Bigg\{\begin{array}{cccc}
  \text{Server I} & \text{Server II} & \text{Server III} & \text{Server IV}\\\hline
  a_{6\lambda+31}+c_{6\lambda+31} & a_{6\lambda+31}+c_{6\lambda+31} & a_{6\lambda+32}+c_{6\lambda+32} & a_{6\lambda+32}+c_{6\lambda+32} \\
  a_{6\lambda+33}+c_{6\lambda+33} & a_{6\lambda+34}+c_{6\lambda+34} & a_{6\lambda+33}+c_{6\lambda+33} & a_{6\lambda+34}+c_{6\lambda+34} \\
  a_{6\lambda+35}+c_{6\lambda+35} & a_{6\lambda+36}+c_{6\lambda+36} & a_{6\lambda+36}+c_{6\lambda+36} & a_{6\lambda+35}+c_{6\lambda+35} \\
\end{array}\Bigg\}\text{ for $\lambda\in\{1,2,3,4,5\}$,}$$
and finally
$$\Lambda^X_{\eta}:\Bigg\{\begin{array}{cccc}
  \text{Server I} & \text{Server II} & \text{Server III} & \text{Server IV}\\\hline
x_{6\eta+61} & x_{6\eta+61} & x_{6\eta+62} & x_{6\eta+62} \\
x_{6\eta+63} & x_{6\eta+64} & x_{6\eta+63} & x_{6\eta+64} \\
x_{6\eta+65} & x_{6\eta+66} & x_{6\eta+66} & x_{6\eta+65} \\
\end{array}\Bigg\}\text{ for $(x,X)\in\{(a,U),(b,V),(c,W)\}$ and $\eta\in\{1,\cdots,25\}$.}$$

The remaining task is then to determine the atoms. Independently choose three random matrices $S_1,S_2,S_3\in\mathbb{F}_q^{216\times216}$, uniformly from all the $216\times216$ full rank matrices over $\mathbb{F}_q$. The atoms $a_{[1:216]}$ are built just by setting $a_{[1:216]}=S_1U$. In Step 2 we have a $(36,30)$-MDS code and the transpose of its generator matrix is denoted as $\text{MDS}_{36\times30}$. Select the first 30 rows of $S_2$, denoted as $S_2[(1:30),:]$. Then $b_{[1:36]}$ are built by setting $b_{[1:36]}=\text{MDS}_{36\times30}S_2[(1:30),:]V$. Similarly $c_{[1:36]}=\text{MDS}_{36\times30}S_3[(1:30),:]W$. This is exactly what we do in Step 7 by choosing $\mathcal{D}=\{V,W\}$. We pause here to explain the motivations of the atoms defined so far. Notice that $\{b_i+c_i:7\le i \le 36\}$ could be retrieved since each appears as a query in two different servers. Then due to the property of $\text{MDS}_{36\times30}$, we can solve $\{b_i+c_i:1\le i \le 6\}$ from $\{b_i+c_i:7\le i \le 36\}$. Therefore the interferences can be eliminated in the block $\Lambda^{UVW}$ and thus $a_{[1:6]}$ could be retrieved. This is exactly how we make use of the side information provided by $\{\Lambda^{VW}_{\lambda}:1\le\lambda\le5\}$ to retrieve some messages of the desired file $U$ hidden in $\Lambda^{UVW}$.

The motivations of the other atoms follow the same idea, i.e., we want to make use of the side information provided by $\{\Lambda^{V}_{\eta}:1\le\eta\le25\}$ (respectively, provided by $\{\Lambda^{W}_{\eta}:1\le\eta\le25\}$) to retrieve some messages of the desired file $U$ hidden in $\{\Lambda^{UV}_{\lambda}:1\le\lambda\le5\}$ (respectively, hidden in $\{\Lambda^{UW}_{\lambda}:1\le\lambda\le5\}$). This is done in separate parallel steps. Divide the query structure $\{\Lambda^{UV}_{\lambda}:1\le \lambda \le 5\},\{\Lambda^{UW}_{\lambda}:1\le \lambda \le 5\},\{\Lambda^X_{\eta}:X\in\{V,W\},1\le \eta \le 25\}$ into the following separate groups:
$$\Gamma^X_{\lambda}=\big\{\Lambda^{UX}_{\lambda},\{\Lambda^X_{\eta}:5\lambda-4\le \eta \le 5\lambda\} \big\},~X\in\{V,W\},~1\le \lambda \le 5.$$
In each $\Gamma^X_{\lambda}$ we can select the atoms as
$$b_{[6\lambda+31:6\lambda+36]\bigcup[30\lambda+37:30\lambda+66]}=\text{MDS}_{36\times30}S_2[(30\lambda+1:30\lambda+30),:]V,$$
or similarly
$$c_{[6\lambda+31:6\lambda+36]\bigcup[30\lambda+37:30\lambda+66]}=\text{MDS}_{36\times30}S_3[(30\lambda+1:30\lambda+30),:]W.$$

Retrieving the file $U$ is equivalent to retrieving $a_{[1:216]}$ since $S_1$ is a full rank matrix. Retrieving $a_{[67:216]}$ is straightforward from the blocks labelled by $U$ and we have explained how to retrieve $a_{[1:6]}$. Then in each $\Gamma^V_{\lambda}$ or $\Gamma^W_{\lambda}$ we have the atoms $a_{[6\lambda+1:6\lambda+6]}$ or $a_{[6\lambda+31:6\lambda+36]}$ accompanied by the interferences $b_{[6\lambda+31:6\lambda+36]}$ or $c_{[6\lambda+31:6\lambda+36]}$. Due to the MDS property we can solve these interferences via $b_{[30\lambda+37:30\lambda+66]}$ or $c_{[30\lambda+37:30\lambda+66]}$. Once eliminating these interferences, we are able to retrieve $a_{[7:66]}$.

The scheme is private against any two colluding servers. The query structure is completely symmetric with respect to all the files. From the perspective of any two colluding servers, altogether the number of atoms towards each of the three files is $125+25\times2+5=180$. Extract the coefficients of each atom as a vector in $\mathbb{F}_q^{216}$. Recall how we select the random matrices $S_1$, $S_2$, $S_3$ and the $(36,30)$-MDS code. It turns out that the 180 vectors with respect to each file form a random subspace of dimension 180 in $\mathbb{F}_q^{216}$, so the two servers cannot tell any difference among the atoms towards different files and thus the identity of the retrieved file is disguised.

The rate of the scheme above is then $\frac{216\times2}{12\times25\times3+12\times5\times3+12}=\frac{36}{91}$.

\subsection{Analysis of the general model}

To verify the correctness and privacy of the general scheme and to compute its rate as shown in Theorem \ref{OriginalRate}, please refer to \cite{Zhangnew}. Here we briefly analyze some key ingredients in the scheme, which will shed light on how to adjust the scheme for other variant models.

First, we should guarantee the correctness of the scheme, i.e., the user should succeed in retrieving his desired file. In the scheme above, all the atoms towards the desired file are independent and the redundancy of the atoms towards undesired files only exists within groups. In such a way we can fully use the side information and retrieve as many bits of the desired files as possible. However, in the model with robust or Byzantine servers, since some feedbacks are either erased or in error, then it is natural to bring in more redundancy among the atoms for both the desired and undesired files. Moreover, note that a robust or Byzantine server not only affects its own feedbacks, it also results in a waste of some other feedbacks from well-functioned servers. For example, if Server I in the previous example is robust, then the user does not know the coded bit $(a_{1}+b_{1}+c_{1})\mathbf{g_1}$. As a result, the coded bit $(a_{1}+b_{1}+c_{1})\mathbf{g_2}$ from Server II will be of no use, but it still accounts for one bit in the total download size.

Second, we should guarantee the privacy of the scheme. We should carefully adjust the ratio of the number of different blocks and this is exactly what the parameters $\alpha$ and $\beta$ computed in Step 2 are aimed for. What we do in the general scheme is to guarantee that from the perspective of any $T$ colluding servers, the subspaces derived from the atoms towards each file are indistinguishable. For the variant models, since we have to bring in more redundancy among the atoms for each file, then we should readjust the ratio correspondingly to guarantee the privacy.

Besides these two major aspects mainly concentrated on modifying the atoms and ratio of blocks, the scheme could be modified through other ways, such as modifying the assisting array or modifying the query structure. We will encounter all kinds of such modifications by analyzing the variant models case-by-case later.

\section{PIR with robust servers} \label{SecRobust}

In this section, we deal with the first variant model, PIR with robust servers. In this model we have a database with $N$ servers storing $M$ files. Each file is stored independently via an arbitrary $(N,K)$-MDS code. A set of $S$ servers do not respond and are therefore called robust servers. A user wants to retrieve a certain file from the database. At first he does not know which set of $S$ servers are robust and thus he makes queries to every server. Of course he will then know which servers are robust since he does not receive feedbacks from these servers. Any set of $T$ servers may collude in attacking on the privacy of the user. Note that a robust server, although providing no feedbacks, may also participate in attacking the privacy of the user. This model can be regarded as an erasure-correcting PIR model.

\subsection{PIR scheme with robust servers}

Constructing the PIR scheme with robust servers contains the following steps.

$\bullet$ Step 1: Let $\alpha$ and $\beta$ be the smallest positive integers satisfying
\begin{equation} \label{equationRobust}
\alpha{{N-S}\choose{K}}=(\alpha+\beta)\Bigg({{N}\choose{K}}-{{N-T}\choose{K}}\Bigg).
\end{equation}
Assume the existence of an $\big((\alpha+\beta){{N}\choose{K}},\alpha{{N-S}\choose{K}}\big)$-MDS code. The transpose of its generator matrix is denoted as $\text{MDS}_{(\alpha+\beta){{N}\choose{K}}\times\alpha{{N-S}\choose{K}}}$. Assume the existence of an $\big({{N}\choose{K}},{{N-S}\choose{K}}\big)$-MDS code. The transpose of its generator matrix is denoted as $\text{MDS}_{{{N}\choose{K}}\times{{N-S}\choose{K}}}$.

$\bullet$ Step 2: Let $L={{N-S}\choose{K}}(\alpha+\beta)^{M-1}$. Independently choose $M$ random matrices, $S_1,\dots,S_M$, uniformly from all the $L\times L$ full rank matrices over $\mathbb{F}_q$.

$\bullet$ Step 3: Build an {\it assisting array} of size ${{N-1}\choose{K-1}}\times N$ consisting of ${{N}\choose{K}}$ symbols. Each symbol appears $K$ times and every $K$ columns share a common symbol.

$\bullet$ Step 4: \emph{[Building basic blocks of the query structure]} The query structure is divided into several blocks, where each block is labelled by a subset of files $\mathcal{D}\subseteq\{W^{[1]},W^{[2]},\dots,W^{[M]}\}$. In a block labelled by $\mathcal{D}$, every query is a mixture of $|\mathcal{D}|$ atoms related to the files in $\mathcal{D}$. We set each block in an ``isomorphic" form with the assisting array, i.e., every $K$ servers share a common query. We further call a block labelled by $\mathcal{D}$ a {\it $d$-block} if $|\mathcal{D}|=d$. For any $\mathcal{D}\subseteq\{W^{[1]},W^{[2]},\dots,W^{[M]}\}$, $|\mathcal{D}|=d$, we require that the number of $d$-blocks labelled by $\mathcal{D}$ is $\alpha^{M-d}\beta^{d-1}$.

$\bullet$ Step 5: \emph{[Constructing the atoms for the desired file in each block]} As indicated by the assisting array, in a block labelled by $\mathcal{D}$ with $W^{[1]}\in\mathcal{D}$, we have ${N\choose K}$ atoms towards the desired file. Let these atoms be built by $\text{MDS}_{{{N}\choose{K}}\times{{N-S}\choose{K}}}S'_1W^{[1]}$, where $S'_1$ denotes some ${{N-S}\choose{K}}$ rows in the matrix $S_1$. Repeat this process for all the blocks labelled by $\mathcal{D}$ with $W^{[1]}\in\mathcal{D}$, totally the number of such blocks is

\begin{equation}
  \sum_{|\mathcal{D}|=1}^{M} \alpha^{M-|\mathcal{D}|}\beta^{|\mathcal{D}|-1}{{M-1}\choose{|\mathcal{D}|-1}}=(\alpha+\beta)^{M-1}.
\end{equation}

We have to ensure that the rows we select from $S_1$ are non-intersecting. This can be satisfied since $(\alpha+\beta)^{M-1}{{N-S}\choose{K}}=L$.

$\bullet$ Step 6: \emph{[Dividing the query structure into groups]} Now let $\mathcal{D}$ denote a nonempty set of files not containing the desired file $W^{[1]}$. There are totally $\alpha^{M-|\mathcal{D}|}\beta^{|\mathcal{D}|-1}$ blocks labelled by $\mathcal{D}$ and $\alpha^{M-|\mathcal{D}|-1}\beta^{|\mathcal{D}|}$ blocks labelled by $\mathcal{D}\bigcup\{W^{[1]}\}$. Partition these blocks into $\alpha^{M-|\mathcal{D}|-1}\beta^{|\mathcal{D}|-1}$ groups, denoted by $\Gamma^{\mathcal{D}}_{\lambda}$, $1\le\lambda\le \alpha^{M-|\mathcal{D}|-1}\beta^{|\mathcal{D}|-1}$, where each group consists of $\alpha$ blocks labelled by $\mathcal{D}$ and $\beta$ blocks labelled by $\mathcal{D}\bigcup\{W^{[1]}\}$. Repeat this process for every nonempty set $\mathcal{D}$ not containing $W^{[1]}$ and thus the whole query structure is partitioned into groups.

$\bullet$ Step 7: \emph{[Constructing the atoms for undesired files in each group]} For each group constructed with respect to $\mathcal{D}$ and any file $W^{[m]}\in\mathcal{D}$, the number of atoms towards $W^{[m]}$ within the blocks in this group is $(\alpha+\beta){{N}\choose{K}}$, among which $\alpha{{N}\choose{K}}$ atoms appear in blocks labelled by $\mathcal{D}$ and the other $\beta{{N}\choose{K}}$ atoms appear as interferences in blocks labelled by $\mathcal{D}\bigcup\{W^{[1]}\}$. Let these $(\alpha+\beta){{N}\choose{K}}$ atoms for $W^{[m]}$ be built by $\text{MDS}_{(\alpha+\beta){{N}\choose{K}}\times\alpha{{N-S}\choose{K}}}S'_mW^{[m]}$, where $S'_m$ denotes some $\alpha{{N-S}\choose{K}}$ rows in the matrix $S_m$.

Repeat this process for all the groups. Note that we shall come across the same file $W^{[M]}$ several times. We have ${{M-2}\choose{|\mathcal{D}|-1}}$ choices for a subset $\mathcal{D}$ of size $|\mathcal{D}|$, containing $W^{[M]}$ but without $W^{[1]}$. With respect to any such $\mathcal{D}$ we have $\alpha^{M-|\mathcal{D}|-1}\beta^{|\mathcal{D}|-1}$ groups. So the exact number of times we come across $W^{[M]}$ can be computed as
\begin{equation}
  \sum_{|\mathcal{D}|=1}^{M-1} \alpha^{M-|\mathcal{D}|-1}\beta^{|\mathcal{D}|-1}{{M-2}\choose{|\mathcal{D}|-1}}=(\alpha+\beta)^{M-2}.
\end{equation}

Every time we come across the same file $W^{[M]}$, we have to ensure that the rows we select from $S_m$ are non-intersecting. This can be satisfied since $(\alpha+\beta)^{M-2}\alpha{{N-S}\choose{K}}\le L$.

The construction of the PIR scheme with robust servers is finished. Comparing the scheme with the prototype, one can see that the main adaptation is changing the redundancy in atoms and then adjusting the ratio of different blocks (i.e. determining the value of parameters $\alpha$ and $\beta$).

\subsection{An example: $N=6$, $S=1$, $K=2$, $T=2$ and $M=2$}

The parameters throughout the construction can be calculated as $\alpha=9$, $\beta=1$ and $L=100$. The assisting array could be of the form
$$\Bigg\{\begin{array}{cccccc}
  1 & 1 & 2 & 3 & 4 & 5\\
  2 & 6 & 6 & 7 & 8 & 9\\
  3 & 7 & 10 & 10 & 11 & 12\\
  4 & 8 & 11 & 13 & 13 & 14\\
  5 & 9 & 12 & 14 & 15 & 15
\end{array}\Bigg\}.$$ Now we have two files $U$ and $V$. Let each file be of length $200$ and represented in a matrix of size $100\times2$, i.e.,
\begin{equation}
  U=\Bigg(
                  \begin{array}{c}
                    \mathbf{u}_1 \\
                    \vdots \\
                    \mathbf{u}_{100} \\
                  \end{array}
                \Bigg)
  \hspace{3em}
  V=\Bigg(
                  \begin{array}{c}
                    \mathbf{v}_1 \\
                    \vdots \\
                    \mathbf{v}_{100} \\
                  \end{array}
                \Bigg)
\end{equation}
where $\mathbf{u}_i,\mathbf{v}_i\in\mathbb{F}_q^{2}$, $1\le i \le 100$. Here $\mathbb{F}_q$ is a sufficiently large finite field. Let $U$ be the desired file. We shall construct two lists of atoms $a_{[1:150]}$ and $b_{[1:150]}$ corresponding to each file accordingly. Following Step 4, the query structure is divided into several blocks as follows. In each block we set the queries in an ``isomorphic" form with the assisting array.

$$\Lambda^{UV}:\Bigg\{\begin{array}{cccccc}
  \text{Server I} & \text{Server II} & \text{Server III} & \text{Server IV} & \text{Server V} & \text{Server VI}\\\hline
  a_{1}+b_{1} & a_{1}+b_{1} & a_{2}+b_{2} & a_{3}+b_{3} & a_{4}+b_{4} & a_{5}+b_{5}\\
  a_{2}+b_{2} & a_{6}+b_{6} & a_{6}+b_{6} & a_{7}+b_{7} & a_{8}+b_{8} & a_{9}+b_{9}\\
  a_{3}+b_{3} & a_{7}+b_{7} & a_{10}+b_{10} & a_{10}+b_{10} & a_{11}+b_{11} & a_{12}+b_{12}\\
  a_{4}+b_{4} & a_{8}+b_{8} & a_{11}+b_{11} & a_{13}+b_{13} & a_{13}+b_{13} & a_{14}+b_{14}\\
  a_{5}+b_{5} & a_{9}+b_{9} & a_{12}+b_{12} & a_{14}+b_{14} & a_{15}+b_{15} & a_{15}+b_{15}
\end{array}\Bigg\}\text{,}$$

$$\Lambda^X_{\eta}:\Bigg\{\begin{array}{cccccc}
  \text{Server I} & \text{Server II} & \text{Server III} & \text{Server IV} & \text{Server V} & \text{Server VI}\\\hline
  x_{15\eta+1} & x_{15\eta+1} & x_{15\eta+2} & x_{15\eta+3} & x_{15\eta+4} & x_{15\eta+5}\\
  x_{15\eta+2} & x_{15\eta+6} & x_{15\eta+6} & x_{15\eta+7} & x_{15\eta+8} & x_{15\eta+9}\\
  x_{15\eta+3} & x_{15\eta+7} & x_{15\eta+10} & x_{15\eta+10} & x_{15\eta+11} & x_{15\eta+12}\\
  x_{15\eta+4} & x_{15\eta+8} & x_{15\eta+11} & x_{15\eta+13} & x_{15\eta+13} & x_{15\eta+14}\\
  x_{15\eta+5} & x_{15\eta+9} & x_{15\eta+12} & x_{15\eta+14} & x_{15\eta+15} & x_{15\eta+15}
\end{array}\Bigg\}\text{ for $(x,X)\in\{(a,U),(b,V)\}$ and $\eta\in\{1,\cdots,9\}$.}$$

The remaining task is then to determine the atoms. In Step 2 we have a $(150,90)$-MDS code and the transpose of its generator matrix is denoted as $\text{MDS}_{150\times90}$. We also have a $(15,10)$-MDS code and the transpose of its generator matrix is denoted as $\text{MDS}_{15\times10}$. Independently choose two random matrices $S_1,S_2\in\mathbb{F}_q^{100\times100}$, uniformly from all the $100\times100$ full rank matrices over $\mathbb{F}_q$. By Step 5, the atoms $a_{[15\eta+1:15\eta+15]}$, $0\le \eta \le 9$ are built by  $\text{MDS}_{15\times10}S_1[(10\eta+1:10\eta+10),:]U$, where $S_1[(10\eta+1:10\eta+10),:]$ means the ten rows indexed from $10\eta+1$ to $10\eta+10$ from $S_1$. By Step 7, the atoms $b_{[1:150]}$ are built by $\text{MDS}_{150\times90}S_2[(1:90),:]V$.

Without loss of generality, suppose Server I is the robust server. Then the user can retrieve $\{a_i+b_i:6\le i \le 15\}$, $\{x_{15\eta+i}:x\in\{a,b\},1\le \eta \le 9,6\le i \le 15\}$. By the property of $\text{MDS}_{150\times90}$, the user can use the 90 known atoms towards $V$ to solve the remaining atoms towards $V$. Particularly, he can solve $\{b_i:6\le i \le 15\}$. Eliminating the interferences, the user now retrieves $\{a_i:6\le i \le 15\}$. Now that the user knows $\{a_{15\eta+i}:0\le \eta \le 9, 6\le i \le 15\}$. By the property of $\text{MDS}_{15\times10}$ and the singularity of $S_1$, the user retrieves the whole file $U$.

The scheme is private against any two colluding servers. From the perspective of any two colluding servers, the number of atoms towards each file is $90$. Extract the coefficients of each atom as a vector in $\mathbb{F}_q^{100}$. Recall how we select the random matrices $S_1$, $S_2$ and those two MDS codes. It turns out that the 90 vectors with respect to each file form a random subspace of dimension 90 in $\mathbb{F}_q^{100}$, so the two servers cannot tell any difference between the atoms towards different files and thus the identity of the retrieved file is disguised.

The rate of the scheme above is then $\frac{100\times2}{5\times5\times19}=\frac{8}{19}$. Note that the denominator is not $6\times5\times19$ since the robust server will produce no download cost at all.

\subsection{Analysis of the PIR scheme with robust servers}

Each query in the scheme appears in $K$ distinct servers. If all these $K$ servers are not robust, then this query can be successfully retrieved. For each group constructed with respect to $\mathcal{D}$, $\alpha{{N-S}\choose{K}}$ queries in the blocks labelled by $\mathcal{D}$ and $\beta{{N-S}\choose{K}}$ queries in the blocks labelled by $\mathcal{D}\bigcup\{W^{[1]}\}$ could be retrieved. Based on the property of the $\big((\alpha+\beta){{N}\choose{K}},\alpha{{N-S}\choose{K}}\big)$-MDS code, these $\alpha{{N-S}\choose{K}}$ queries in the blocks labelled by $\mathcal{D}$ will help us eliminate the interferences of the $\beta{{N-S}\choose{K}}$ queries in the blocks labelled by $\mathcal{F}\bigcup\{W^{[1]}\}$, and thus the $\beta{{N-S}\choose{K}}$ atoms towards $W^{[1]}$ are retrieved. As a result, in each block related to $W^{[1]}$, ${{N-S}\choose{K}}$ out of the ${{N}\choose{K}}$ atoms towards $W^{[1]}$ are successfully retrieved. By the property of $\text{MDS}_{{{N}\choose{K}}\times{{N-S}\choose{K}}}$ and the singularity of $S_1$, the desired file is successfully retrieved.

The scheme is private against any $T$ colluding servers. From the perspective of any $T$ colluding servers, they shall receive $(\alpha+\beta)^{M-1}({N\choose K}-{{N-T}\choose{K}})$ atoms towards each file. Extract the coefficients of each atom as a vector in $\mathbb{F}_q^{L}$. Recall how we build the atoms by using the full rank matrices $S_1,\dots,S_m$ and the two MDS codes. One can see that the vectors of all the atoms towards any file will form a random subspace of dimension $(\alpha+\beta)^{M-1}({N\choose K}-{{N-T}\choose{K}})$ in $\mathbb{F}_q^{L}$. So these $T$ colluding servers cannot tell any difference among the atoms towards different files and thus the identity of the retrieved file is disguised.

Finally we compute the rate of the PIR scheme with robust servers. The retrieved file is of length $LK$. The download cost in any block is $K{N\choose K}\frac{N-S}{N}$. So the rate can be computed as
$$\frac{LK}{K{N\choose K}\frac{N-S}{N}\sum_{|\mathcal{D}|=1}^{M} \alpha^{M-|\mathcal{D}|}\beta^{|\mathcal{D}|-1}{{M}\choose{|\mathcal{D}|}}}=\frac{L}{{N\choose K}\frac{N-S}{N}\beta^{-1}\big((\alpha+\beta)^M-\alpha^M\big)}=\frac{{{N-S-1}\choose{K-1}}}{{{N-1}\choose{K-1}}}\cdot\frac{1}{1+\frac{\alpha}{\alpha+\beta}+\cdots+(\frac{\alpha}{\alpha+\beta})^{M-1}}.$$

A final remark is that our scheme only works for the case $T+K\le N$ and ${{N-S}\choose K}>{{N}\choose K}-{{N-T}\choose K}$ since otherwise the equality (\ref{equationRobust}) does not make sense. To sum up, we have the following theorem.

\begin{theorem}
  When $T+K\le N$ and ${{N-S}\choose K}>{{N}\choose K}-{{N-T}\choose K}$, there exists an $(N,K,T;M)$-PIR scheme with $S$ robust servers whose rate is $\frac{{{N-S-1}\choose{K-1}}}{{{N-1}\choose{K-1}}}(1+R+R^2+\cdots+R^{M-1})^{-1}$, where $R=\frac{{N\choose K}-{{N-T}\choose K}}{{{N-S}\choose K}}$.
\end{theorem}

Particularly, for the degenerating case $K=1$, the PIR rate of the scheme is $(1+R+R^2+\cdots+R^{M-1})^{-1}$ with $R=\frac{T}{N-S}$, which is exactly the result presented in \cite{SunColluded}.

\section{PIR with Byzantine servers} \label{SecByzantine}

In this section, we deal with the second variant model, PIR with Byzantine servers. In this model we have a database with $N$ servers storing $M$ files. Each file is stored independently via an arbitrary $(N,K)$-MDS code. A set of $B$ servers are called Byzantine servers. They will produce incorrect feedbacks due to their outdated contents or even adversary motivations. A user wants to retrieve a certain file from the database and he does not know which set of $B$ servers are Byzantine. Any set of $T$ servers may collude in attacking on the privacy of the user. We should differentiate the actions of $T$ servers colluding in attacking on the privacy and $B$ servers coordinating in introducing errors. No specific relation between these two actions are assumed. This model can be regarded as an error-correcting PIR model. Recall the fact from classic coding theory: the maximum number of errors that can be corrected by a linear code with minimum distance $d$ is $\tau\le\lfloor\frac{d-1}{2}\rfloor$. As for an $[n,k,d]$-linear MDS code, $\tau\le\lfloor\frac{d-1}{2}\rfloor=\lfloor\frac{n-k}{2}\rfloor$.

\subsection{PIR scheme with Byzantine servers}

Constructing the PIR scheme with robust servers contains the following steps.

$\bullet$ Step 1: Let $\alpha$ and $\beta$ be the smallest positive integers satisfying
\begin{equation} \label{equationByzantine}
\alpha\Bigg(2{{N-B}\choose{K}}-{{N}\choose{K}}\Bigg)=(\alpha+\beta)\Bigg({{N}\choose{K}}-{{N-T}\choose{K}}\Bigg).
\end{equation}
Assume the existence of an $\big((\alpha+\beta){{N}\choose{K}},\alpha(2{{N-B}\choose{K}}-{{N}\choose{K}})\big)$-MDS code. The transpose of its generator matrix is denoted as $\text{MDS}_{(\alpha+\beta){{N}\choose{K}}\times\alpha\big(2{{N-B}\choose{K}}-{{N}\choose{K}}\big)}$. Assume the existence of an $\big({{N}\choose{K}},2{{N-B}\choose{K}}-{{N}\choose{K}}\big)$-MDS code. The transpose of its generator matrix is denoted as $\text{MDS}_{{{N}\choose{K}}\times\big(2{{N-B}\choose{K}}-{{N}\choose{K}}\big)}$.

$\bullet$ Step 2: Let $L=\big(2{{N-B}\choose{K}}-{{N}\choose{K}}\big)(\alpha+\beta)^{M-1}$. Independently choose $M$ random matrices, $S_1,\dots,S_M$, uniformly from all the $L\times L$ full rank matrices over $\mathbb{F}_q$.

$\bullet$ Step 3: Build an {\it assisting array} of size ${{N-1}\choose{K-1}}\times N$ consisting of ${{N}\choose{K}}$ symbols. Each symbol appears $K$ times and every $K$ columns share a common symbol.

$\bullet$ Step 4: \emph{[Building basic blocks of the query structure]} The query structure is divided into several blocks, where each block is labelled by a subset of files $\mathcal{D}\subseteq\{W^{[1]},W^{[2]},\dots,W^{[M]}\}$. In a block labelled by $\mathcal{D}$, every query is a mixture of $|\mathcal{D}|$ atoms related to the files in $\mathcal{D}$. We set each block in an ``isomorphic" form with the assisting array, i.e., every $K$ servers share a common query. We further call a block labelled by $\mathcal{D}$ a {\it $d$-block} if $|\mathcal{D}|=d$. For any $\mathcal{D}\subseteq\{W^{[1]},W^{[2]},\dots,W^{[M]}\}$, $|\mathcal{D}|=d$, we require that the number of $d$-blocks labelled by $\mathcal{D}$ is $\alpha^{M-d}\beta^{d-1}$.

$\bullet$ Step 5: \emph{[Constructing the atoms for the desired file in each block]} As indicated by the assisting array, in a block labelled by $\mathcal{D}$ with $W^{[1]}\in\mathcal{D}$, we have ${N\choose K}$ atoms towards the desired file. Let these atoms be built by $\text{MDS}_{{{N}\choose{K}}\times\big(2{{N-B}\choose{K}}-{{N}\choose{K}}\big)}S'_1W^{[1]}$, where $S'_1$ denotes some $2{{N-B}\choose{K}}-{{N}\choose{K}}$ rows in the matrix $S_1$. Repeat this process for all the blocks labelled by $\mathcal{D}$ with $W^{[1]}\in\mathcal{D}$, totally the number of such blocks is

\begin{equation}
  \sum_{|\mathcal{D}|=1}^{M} \alpha^{M-|\mathcal{D}|}\beta^{|\mathcal{D}|-1}{{M-1}\choose{|\mathcal{D}|-1}}=(\alpha+\beta)^{M-1}.
\end{equation}

We have to ensure that the rows we select from $S_1$ are non-intersecting. This can be satisfied since $(\alpha+\beta)^{M-1}\times\big(2{{N-B}\choose{K}}-{{N}\choose{K}}\big) = L$.

$\bullet$ Step 6: \emph{[Dividing the query structure into groups]} Now let $\mathcal{D}$ denote a nonempty set of files not containing the desired file $W^{[1]}$. There are totally $\alpha^{M-|\mathcal{D}|}\beta^{|\mathcal{D}|-1}$ blocks labelled by $\mathcal{D}$ and $\alpha^{M-|\mathcal{D}|-1}\beta^{|\mathcal{D}|}$ blocks labelled by $\mathcal{D}\bigcup\{W^{[1]}\}$. Partition these blocks into $\alpha^{M-|\mathcal{D}|-1}\beta^{|\mathcal{D}|-1}$ groups, denoted by $\Gamma^{\mathcal{D}}_{\lambda}$, $1\le\lambda\le \alpha^{M-|\mathcal{D}|-1}\beta^{|\mathcal{D}|-1}$, where each group consists of $\alpha$ blocks labelled by $\mathcal{D}$ and $\beta$ blocks labelled by $\mathcal{D}\bigcup\{W^{[1]}\}$. Repeat this process for every nonempty set $\mathcal{D}$ not containing $W^{[1]}$ and thus the whole query structure is partitioned into groups.

$\bullet$ Step 7: \emph{[Constructing the atoms for undesired files in each group]} For each group constructed with respect to $\mathcal{D}$ and any file $W^{[m]}\in\mathcal{D}$, the number of atoms towards $W^{[m]}$ within the blocks in this group is $(\alpha+\beta){{N}\choose{K}}$, among which $\alpha{{N}\choose{K}}$ atoms appear in blocks labelled by $\mathcal{D}$ and the other $\beta{{N}\choose{K}}$ atoms appear as interferences in blocks labelled by $\mathcal{D}\bigcup\{W^{[1]}\}$. Let these $(\alpha+\beta){{N}\choose{K}}$ atoms for $W^{[m]}$ be built by $\text{MDS}_{(\alpha+\beta){{N}\choose{K}}\times\alpha\big(2{{N-B}\choose{K}}-{{N}\choose{K}}\big)}S'_mW^{[m]}$, where $S'_m$ denotes some $\alpha\big(2{{N-B}\choose{K}}-{{N}\choose{K}}\big)$ rows in the matrix $S_m$.

Repeat this process for all the groups. Note that we shall come across the same file $W^{[M]}$ several times. We have ${{M-2}\choose{|\mathcal{D}|-1}}$ choices for a subset $\mathcal{D}$ of size $|\mathcal{D}|$, containing $W^{[M]}$ but without $W^{[1]}$. With respect to any such $\mathcal{D}$ we have $\alpha^{M-|\mathcal{D}|-1}\beta^{|\mathcal{D}|-1}$ groups. So the exact number of times we come across $W^{[M]}$ can be computed as
\begin{equation}
  \sum_{|\mathcal{D}|=1}^{M-1} \alpha^{M-|\mathcal{D}|-1}\beta^{|\mathcal{D}|-1}{{M-2}\choose{|\mathcal{D}|-1}}=(\alpha+\beta)^{M-2}.
\end{equation}

Every time we come across the same file $W^{[M]}$, we have to ensure that the rows we select from $S_m$ are non-intersecting. This can be satisfied since
\begin{equation} \label{InequalityByzantine2}
  (\alpha+\beta)^{M-2}\alpha\Bigg(2{{N-B}\choose{K}}-{{N}\choose{K}}\Bigg)\le L.
\end{equation}

The construction of the PIR scheme with Byzantine servers is finished. Comparing the scheme with the prototype, one can see that the main adaptation is changing the redundancy in atoms and then adjusting the ratio of different blocks (i.e. determining the value of parameters $\alpha$ and $\beta$).

\subsection{An example: $N=8$, $B=1$, $K=2$, $T=2$ and $M=2$}

The parameters throughout the construction can be calculated as $\alpha=13$, $\beta=1$ and $L=196$. The assisting array could be of the form
$$\Bigg\{\begin{array}{cccccccc}
  1 & 1 & 2 & 3 & 4 & 5 & 6 & 7 \\
  2 & 8 & 8 & 9 & 10 & 11 & 12 & 13 \\
  3 & 9 & 14 & 14 & 15 & 16 & 17 & 18 \\
  4 & 10 & 15 & 19 & 19 & 20 & 21 & 22 \\
  5 & 11 & 16 & 20 & 23 & 23 & 24 & 25 \\
  6 & 12 & 17 & 21 & 24 & 26 & 26 & 27 \\
  7 & 13 & 18 & 22 & 25 & 27 & 28 & 28 \\
\end{array}\Bigg\}.$$

Now we have two files $U$ and $V$. Let each file be of length $392$ and represented in a matrix of size $196\times2$, i.e.,
\begin{equation}
  U=\Bigg(
                  \begin{array}{c}
                    \mathbf{u}_1 \\
                    \vdots \\
                    \mathbf{u}_{196} \\
                  \end{array}
                \Bigg)
  \hspace{3em}
  V=\Bigg(
                  \begin{array}{c}
                    \mathbf{v}_1 \\
                    \vdots \\
                    \mathbf{v}_{196} \\
                  \end{array}
                \Bigg)
\end{equation}
where $\mathbf{u}_i,\mathbf{v}_i\in\mathbb{F}_q^{2}$, $1\le i \le 196$. Here $\mathbb{F}_q$ is a sufficiently large finite field. Let $U$ be the desired file. We shall construct two lists of atoms $a_{[1:392]}$ and $b_{[1:392]}$ corresponding to each file accordingly. Following Step 4, the query structure is divided into several blocks as follows. In each block we set the queries in an ``isomorphic" form with the assisting array.

$$\Lambda^{UV}:\Bigg\{\begin{array}{cccccccc}
  \text{Server I} & \text{Server II} & \text{Server III} & \text{Server IV} & \text{Server V} & \text{Server VI} & \text{Server VII} & \text{Server VIII} \\\hline
  a_{1}+b_{1} & a_{1}+b_{1} & a_{2}+b_{2} & a_{3}+b_{3} & a_{4}+b_{4} & a_{5}+b_{5} & a_{6}+b_{6} & a_{7}+b_{7}\\
  a_{2}+b_{2} & a_{8}+b_{8} & a_{8}+b_{8} & a_{9}+b_{9} & a_{10}+b_{10} & a_{11}+b_{11} & a_{12}+b_{12} & a_{13}+b_{13}\\
  a_{3}+b_{3} & a_{9}+b_{9} & a_{14}+b_{14} & a_{14}+b_{14} & a_{15}+b_{15} & a_{16}+b_{16} & a_{17}+b_{17} & a_{18}+b_{18}\\
  a_{4}+b_{4} & a_{10}+b_{10} & a_{15}+b_{15} & a_{19}+b_{19} & a_{19}+b_{19} & a_{20}+b_{20} & a_{21}+b_{21} & a_{22}+b_{22}\\
  a_{5}+b_{5} & a_{11}+b_{11} & a_{16}+b_{16} & a_{20}+b_{20} & a_{23}+b_{23} & a_{23}+b_{23} & a_{24}+b_{24} & a_{25}+b_{25}\\
  a_{6}+b_{6} & a_{12}+b_{12} & a_{17}+b_{17} & a_{21}+b_{21} & a_{24}+b_{24} & a_{26}+b_{26} & a_{26}+b_{26} & a_{27}+b_{27}\\
  a_{7}+b_{7} & a_{13}+b_{13} & a_{18}+b_{18} & a_{22}+b_{22} & a_{25}+b_{25} & a_{27}+b_{27} & a_{28}+b_{28} & a_{28}+b_{28}
\end{array}\Bigg\}\text{,}$$

$$\Lambda^X_{\eta}:\Bigg\{\begin{array}{cccccccc}
  \text{Server I} & \text{Server II} & \text{Server III} & \text{Server IV} & \text{Server V} & \text{Server VI} & \text{Server VII} & \text{Server VIII} \\\hline
  x_{28\eta+1} & x_{28\eta+1} & x_{28\eta+2} & x_{28\eta+3} & x_{28\eta+4} & x_{28\eta+5} & x_{28\eta+6} & x_{28\eta+7}\\
  x_{28\eta+2} & x_{28\eta+8} & x_{28\eta+8} & x_{28\eta+9} & x_{28\eta+10} & x_{28\eta+11} & x_{28\eta+12} & x_{28\eta+13}\\
  x_{28\eta+3} & x_{28\eta+9} & x_{28\eta+14} & x_{28\eta+14} & x_{28\eta+15} & x_{28\eta+16} & x_{28\eta+17} & x_{28\eta+18}\\
  x_{28\eta+4} & x_{28\eta+10} & x_{28\eta+15} & x_{28\eta+19} & x_{28\eta+19} & x_{28\eta+20} & x_{28\eta+21} & x_{28\eta+22}\\
  x_{28\eta+5} & x_{28\eta+11} & x_{28\eta+16} & x_{28\eta+20} & x_{28\eta+23} & x_{28\eta+23} & x_{28\eta+24} & x_{28\eta+25}\\
  x_{28\eta+6} & x_{28\eta+12} & x_{28\eta+17} & x_{28\eta+21} & x_{28\eta+24} & x_{28\eta+26} & x_{28\eta+26} & x_{28\eta+27}\\
  x_{28\eta+7} & x_{28\eta+13} & x_{28\eta+18} & x_{28\eta+22} & x_{28\eta+25} & x_{28\eta+27} & x_{28\eta+28} & x_{28\eta+28}
\end{array}\Bigg\}$$
for $(x,X)\in\{(a,U),(b,V)\}$ and $\eta\in\{1,\cdots,13\}$.

The remaining task is then to determine the atoms. In Step 2 we have a $(392,182)$-MDS code and the transpose of its generator matrix is denoted as $\text{MDS}_{392\times182}$. We also have a $(28,14)$-MDS code and the transpose of its generator matrix is denoted as $\text{MDS}_{28\times14}$. Independently choose two random matrices $S_1,S_2\in\mathbb{F}_q^{196\times196}$, uniformly from all the $196\times196$ full rank matrices over $\mathbb{F}_q$. By Step 5, the atoms $a_{[28\eta+1:28\eta+28]}$, $0\le \eta \le 13$ are built by setting $\text{MDS}_{28\times14}S_1[(14\eta+1:14\eta+14),:]U$, where $S_1[(14\eta+1:14\eta+14),:]$ means the fourteen rows indexed from $14\eta+1$ to $14\eta+14$ from $S_1$. By Step 7, the atoms $b_{[1:392]}$ are built by $\text{MDS}_{392\times182}S_2[(1:182),:]V$.

The user can retrieve all queries since each appears in $k=2$ servers. However, due to the existence of a Byzantine server, there will be errors and the number of errors in $b_{[29:392]}$ is at most $91$ (at most seven errors from each $\Lambda^V_{\eta}$). Due to the property of $\text{MDS}_{392\times182}$, these errors can be corrected and therefore the user retrieves the correct value of $b_{[1:28]}$. Now the user has $a_{[28\eta+1:28\eta+28]}$, $0\le \eta \le 13$. Due to the property of $\text{MDS}_{28\times14}$, each subset $a_{[28\eta+1:28\eta+28]}$ can correct its possible seven (at most) errors and thus the file $U$ is successfully retrieved.

The scheme is private against any two colluding servers. From the perspective of any two colluding servers, the number of atoms towards each file is $182$. Extract the coefficients of each atom as a vector in $\mathbb{F}_q^{196}$. Recall how we select the random matrices $S_1$, $S_2$ and the two MDS codes. It turns out that the 182 vectors with respect to each file form a random subspace of dimension 182 in $\mathbb{F}_q^{196}$, so the two servers cannot tell any difference between the atoms towards different files and thus the identity of the retrieved file is disguised.

The rate of the scheme above is then $\frac{196\times2}{8\times7\times27}=\frac{7}{27}$.

\subsection{Analysis of the PIR scheme with Byzantine servers}

Each query in the scheme appears in $K$ distinct servers and is thus retrieved. However, due to the existence of a Byzantine server, there will be errors. For each group constructed with respect to $\mathcal{D}$, among the $\alpha{{N}\choose{K}}$ queries in the blocks labelled by $\mathcal{D}$ there will be at most $\alpha\big( {N\choose K}-{{N-B}\choose K}\big)$ errors. The errors can be corrected since these $\alpha{{N}\choose{K}}$ queries arise from a punctured MDS code with minimum distance $1+2\alpha\big( {N\choose K}-{{N-B}\choose K}\big)$. Therefore the user can correctly eliminate the interferences in those $\beta{{N-S}\choose{K}}$ queries in the blocks labelled by $\mathcal{D}\bigcup\{W^{[1]}\}$ and retrieve the atoms towards $W^{[1]}$. Now every atom towards $W^{[1]}$ is retrieved. Then by the property of $\text{MDS}_{{{N}\choose{K}}\times\big(2{{N-B}\choose{K}}-{{N}\choose{K}}\big)}$ and the singularity of $S_1$, the desired file is successfully retrieved.

The scheme is private against any $T$ colluding servers. From the perspective of any $T$ colluding servers, they shall receive $(\alpha+\beta)^{M-1}({N\choose K}-{{N-T}\choose{K}})$ atoms towards each file. Extract the coefficients of each atom as a vector in $\mathbb{F}_q^{L}$. Recall how we build the atoms by using the full rank matrices $S_1,\dots,S_m$ and the two MDS codes. One can see that the vectors of all the atoms towards any file will form a random subspace of dimension $(\alpha+\beta)^{M-1}({N\choose K}-{{N-T}\choose{K}})$ in $\mathbb{F}_q^{L}$. So these $T$ colluding servers cannot tell any difference among the atoms towards different files and thus the identity of the retrieved file is disguised.

Finally we compute the rate of the PIR scheme with robust servers. The retrieved file is of length $LK$. The download cost in any block is $K{N\choose K}$. So the rate can be computed as
$$\frac{LK}{K{N\choose K}\sum_{|\mathcal{F}|=1}^{M} \alpha^{M-|\mathcal{F}|}\beta^{|\mathcal{F}|-1}{{M}\choose{|\mathcal{F}|}}}=\frac{L}{{N\choose K}\beta^{-1}\big((\alpha+\beta)^M-\alpha^M\big)}=\frac{2{{N-B}\choose{K}}-{{N}\choose{K}}}{{N\choose K}}\cdot\frac{1}{1+\frac{\alpha}{\alpha+\beta}+\cdots+(\frac{\alpha}{\alpha+\beta})^{M-1}}.$$

A final remark is that our scheme only works for the case $T+K\le N$ and $2{{N-B}\choose{K}}-{{N}\choose{K}}>{{N}\choose K}-{{N-T}\choose K}$ since otherwise the equality (\ref{equationByzantine}) does not make sense. To sum up, we have the following theorem.

\begin{theorem}
  When $T+K\le N$ and $2{{N-B}\choose{K}}-{{N}\choose{K}}>{{N}\choose K}-{{N-T}\choose K}$, there exists an $(N,K,T;M)$-PIR scheme with $B$ robust servers whose rate is $\frac{2{{N-B}\choose{K}}-{{N}\choose{K}}}{{N\choose K}}(1+R+R^2+\cdots+R^{M-1})^{-1}$, where $R=\frac{{N\choose K}-{{N-T}\choose K}}{2{{N-B}\choose{K}}-{{N}\choose{K}}}$.
\end{theorem}

Particularly, for the degenerating case $K=1$, the PIR rate of the scheme is $\frac{N-2B}{N}(1+R+R^2+\cdots+R^{M-1})^{-1}$ with $R=\frac{T}{N-2B}$, which is exactly the result presented in \cite{BanawanByzantine}.

\section{Multi-file PIR} \label{SecMulti}

In this section, we deal with the multi-file PIR problem. Again we have a database with $N$ servers storing $M$ files and each file is stored independently via an arbitrary $(N,K)$-MDS code. Now a user wants to retrieve several files simultaneously. A naive approach to retrieve $P\ge 2$ files simultaneously is to execute $P$ independent single-file PIR schemes. It should be noted that in a scheme designed for retrieving a certain file, the user may have additional knowledge of some other desired files. However, even taking these benefits into consideration, it is still possible to do better than the naive approach. The multi-file model is a little more complex in the sense that we may have different strategies according to whether the ratio $P/M$ is larger or smaller than $1/2$, as shown in \cite{BanawanMultimessage} for the degenerating case $K=T=1$. In this draft we only deal with the case $P\ge\frac{M}{2}$\footnote{We claim that we have also considered the complementary case $P<\frac{M}{2}$ and adapted our PIR scheme. However, the redundancy within atoms towards desired files are very complex and thus the whole tedious scheme is not very satisfactory and thus omitted. A neater scheme is considered for future research.}.

The rest of this section is further divided into two subsections. The first subsection contains the analysis of the upper bound of the multi-file PIR capacity based on information inequalities. The next subsection discusses the multi-file PIR scheme for $P\ge \frac{M}{2}$.

\subsection{Upper bound of the multi-file PIR capacity for the degenerating cases} \label{Upper}

In this subsection we shall analyze the upper bound of the multi-file PIR capacity based on information inequalities. The case $K=T=1$ has been presented in \cite{BanawanMultimessage}. We shall only consider the degenerating cases $K=1$ or $T=1$ here. As for the non-degenerating case (i.e., $K\ge2$ and $T\ge2$), analyzing the upper bound of the capacity is a rather difficult problem, even for the general single-file PIR model. Given the parameters $P,N,K,T$, the minimal download cost is then a function of the number of files $M$, denoted as $\Omega_M$. For simplicity we need the following notations as in \cite{BanawanMultimessage}.
\begin{equation}
  \mathcal{Q}\triangleq\big\{Q^{[\mathcal{P}]}_n:~\mathcal{P}\subset\{1,2,\dots,M\},~|\mathcal{P}|=P,~n\in\{1,2,\dots,N\} \big\}.
\end{equation}
\begin{equation}
  A^{[\mathcal{P}]}_{n_1:n_2}\triangleq\big\{ A^{[\mathcal{P}]}_{n_1},A^{[\mathcal{P}]}_{n_1+1},\dots,A^{[\mathcal{P}]}_{n_2}\big\} \text{~for~} n_1\le n_2, ~n_1,n_2\in\{1,2,\dots,M\}.
\end{equation}

The scheme could be assumed to be symmetric, or else we could build a symmetric scheme via a non-symmetric scheme by replicating all permutations of servers and files. Therefore we may assume that for any two sets of indices of $P$ desired files $\mathcal{P}_1,\mathcal{P}_2$ and for any two servers, say the $n$th and the $m$th server, we have
\begin{equation}
  H(A^{[\mathcal{P}_1]}_n|\mathcal{Q})=H(A^{[\mathcal{P}_2]}_m|\mathcal{Q}).
\end{equation}

The equality above also holds when one considers conditional entropy with prior knowledge of any subset of files, i.e.,
\begin{equation}
  H(A^{[\mathcal{P}_1]}_n|W^{[\mathcal{S}]},\mathcal{Q})=H(A^{[\mathcal{P}_2]}_m|W^{[\mathcal{S}]},\mathcal{Q}) \text{~for any~} \mathcal{S}\subset\{1,2,\dots,M\}.
\end{equation}

The key to the analysis of the multi-file PIR capacity is the following inductive relation.
\begin{align}
  PL&=H(W^{[\mathcal{P}]})\\
    &=H(W^{[\mathcal{P}]}|\mathcal{Q})\\
    &=I(W^{[\mathcal{P}]};A^{[\mathcal{P}]}_{1:N}|\mathcal{Q})+H(W^{[\mathcal{P}]}|A^{[\mathcal{P}]}_{1:N},\mathcal{Q})\\
    &=I(W^{[\mathcal{P}]};A^{[\mathcal{P}]}_{1:N}|\mathcal{Q})\\
    &=H(A^{[\mathcal{P}]}_{1:N}|\mathcal{Q})-H(A^{[\mathcal{P}]}_{1:N}|W^{[\mathcal{P}]},\mathcal{Q})\\
    &\le \sum_{1\le n\le N} H(A^{[\mathcal{P}]}_{n}|\mathcal{Q})-H(A^{[\mathcal{P}]}_{1:N}|W^{[\mathcal{P}]},\mathcal{Q}).
\end{align}

Note that $\sum_{1\le n\le N} H(A^{[\mathcal{P}]}_{n}|\mathcal{Q})$ is exactly the total download cost. So the inductive relation above can be rewritten as
\begin{equation}
  \Omega_M\ge PL+H(A^{[\mathcal{P}]}_{1:N}|W^{[\mathcal{P}]},\mathcal{Q}).
\end{equation}

The remaining trick is how to bound the second part in the inequality above. We are now ready to analyze the capacity for two degenerating cases, i.e., either $T=1$ or $K=1$.

$\bullet$ Case 1: $T=1$.

From the symmetry argument and the fact that any $K$ servers possess independent messages, we have
\begin{equation} \label{Kindependent}
H(A^{[\mathcal{P}]}_{1:N}|W^{[\mathcal{P}]},\mathcal{Q})\ge KH(A^{[\mathcal{P}]}_{1}|W^{[\mathcal{P}]},\mathcal{Q}).
\end{equation}

One way for a server to analyze $\mathcal{P}$ is by analyzing the entropy of the queries it receives (or equivalently, the feedbacks it produces) conditioned on any subset of files. Since each server is not allowed to know any information of the identity of the retrieved files, we must have
\begin{equation}
  H(A^{[\mathcal{P}]}_{1}|W^{[\mathcal{P}]},\mathcal{Q})=H(A^{[\mathcal{P}]}_{1}|W^{[\mathcal{P}']},\mathcal{Q}),~\forall \mathcal{P}'\subset\{1,2,\dots,M\},~|\mathcal{P}'|=P.
\end{equation}

When $M\ge2P$, we may choose the index set $\mathcal{P}'$ to be disjoint from $\mathcal{P}$. Thus, $H(A^{[\mathcal{P}]}_{1}|W^{[\mathcal{P}']},\mathcal{Q})$ characterises the amount of download from a server in a PIR scheme with the same parameters $P,N,K,T$ but a different number of files $M-P$. So we have the following inductive relation.
\begin{equation} \label{Re1}
  \Omega_M\ge PL+\frac{K}{N}\Omega_{M-P},~M\ge2P.
\end{equation}

Plus, when $M<2P$, $H(A^{[\mathcal{P}]}_{1}|W^{[\mathcal{P}]},\mathcal{Q})$ is lower bounded by $\frac{(M-P)L}{N}$, as shown in \cite[Lemma 2]{BanawanMultimessage}. So we have $\Omega_M\ge PL+\frac{K}{N}(M-P)L$, $M<2P$. Recursively using the inductive relation (\ref{Re1}), we have
\begin{equation}
  \Omega_M\ge PL(1+\frac{K}{N}+(\frac{K}{N})^2+\cdots+(\frac{K}{N})^{\lfloor\frac{M}{P}\rfloor-1}) +(\frac{K}{N})^{\lfloor\frac{M}{P}\rfloor}(M-P\lfloor\frac{M}{P}\rfloor)L.
\end{equation}
So the multi-file PIR capacity $C^P$ is upper bounded by
\begin{equation}\label{CapacityK1}
  C^P=\frac{PL}{\Omega_M}\le\Bigg(\frac{1-(\frac{K}{N})^{\lfloor\frac{M}{P}\rfloor}}{1-\frac{K}{N}}+\big(\frac{M}{P}-\lfloor\frac{M}{P}\rfloor\big)\frac{K^{\lfloor\frac{M}{P}\rfloor}}{N^{\lfloor\frac{M}{P}\rfloor}}\Bigg)^{-1}.
\end{equation}
And in particular, when $M\le2P$, we have
\begin{equation} \label{CapacityK2}
  C^P\le\big(1+\frac{K(M-P)}{PN}\big)^{-1}.
\end{equation}

One can check that when $K=1$, the results (\ref{CapacityK1}) and (\ref{CapacityK2}) agree with those in \cite{BanawanMultimessage}.

$\bullet$ Case 2: $K=1$.

Similarly as (\ref{Kindependent}), we have the following claim.
\begin{equation} \label{Tindependent}
H(A^{[\mathcal{P}]}_{1:N}|W^{[\mathcal{P}]},\mathcal{Q})\ge TH(A^{[\mathcal{P}]}_{1}|W^{[\mathcal{P}]},\mathcal{Q}).
\end{equation}

No matter how we design a PIR scheme, we have two kinds of queries. The first kind is a query dependent on the desired files. The second kind is a query independent on the desired files (serving as side information). Each query of the first kind will contribute one bit to the retrieved messages. In an optimal scheme, it is certain that all the queries of the first kind will contribute independent bits, or otherwise the redundancy could be made use of to improve the scheme. From the perspective of any $T$ colluding servers, since they are not allowed to learn any information of the identity of the retrieved files, then the queries of the second kind on these $T$ servers should contribute independent bits as well. Therefore we have the claim (\ref{Tindependent}) above.


Using the analyses above of $H(A^{[\mathcal{P}]}_{1}|W^{[\mathcal{P}]},\mathcal{Q})$ again, we can deduce the following inductive relation.
\begin{equation} \label{Re2}
  \Omega_M\ge PL+\frac{T}{N}\Omega_{M-P},~M\ge2P.
\end{equation}

Plus, when $M<2P$, $H(A^{[\mathcal{P}]}_{1}|W^{[\mathcal{P}]},\mathcal{Q})$ is lower bounded by $\frac{(M-P)L}{N}$, as shown in \cite[Lemma 2]{BanawanMultimessage}. So we have $\Omega_M\ge PL+\frac{T}{N}(M-P)L$, $M<2P$. Recursively using the inductive relation (\ref{Re2}), we have
\begin{equation}
  \Omega_M\ge PL(1+\frac{T}{N}+(\frac{T}{N})^2+\cdots+(\frac{T}{N})^{\lfloor\frac{M}{P}\rfloor-1}) +(\frac{T}{N})^{\lfloor\frac{M}{P}\rfloor}(M-P\lfloor\frac{M}{P}\rfloor)L.
\end{equation}
So the multi-file PIR capacity $C^P$ is upper bounded by
\begin{equation}\label{CapacityT1}
  C^P=\frac{PL}{\Omega_M}\le\Bigg(\frac{1-(\frac{T}{N})^{\lfloor\frac{M}{P}\rfloor}}{1-\frac{T}{N}}+\big(\frac{M}{P}-\lfloor\frac{M}{P}\rfloor\big)\frac{T^{\lfloor\frac{M}{P}\rfloor}}{N^{\lfloor\frac{M}{P}\rfloor}}\Bigg)^{-1}.
\end{equation}
And in particular, when $M\le2P$, we have
\begin{equation} \label{CapacityT2}
  C^P\le\big(1+\frac{T(M-P)}{PN}\big)^{-1}.
\end{equation}

One can check that when $T=1$, the results (\ref{CapacityT1}) and (\ref{CapacityT2}) agree with those in \cite{BanawanMultimessage}.

\subsection{A general multi-file PIR scheme for $P\ge\frac{M}{2}$}\label{SecSchemeCase1}

In this subsection we shall present a general multi-file PIR scheme for $P\ge\frac{M}{2}$, i.e., a user wants to retrieve at least half of all the files.

\subsubsection{Multi-file PIR scheme for $P\ge\frac{M}{2}$}

Constructing the multi-file PIR scheme for $P\ge\frac{M}{2}$ contains the following steps.

$\bullet$ Step 1: Let $\alpha$ and $\beta$ be the smallest positive integers satisfying
\begin{equation} \label{equationmulti1}
\alpha{{N}\choose{K}}=(\alpha+\beta)\Bigg({{N}\choose{K}}-{{N-T}\choose{K}}\Bigg).
\end{equation}
Assume the existence of an $\big((\alpha+\beta){{N}\choose{K}},\alpha{{N}\choose{K}}\big)$-MDS code. The transpose of its generator matrix is denoted as $\text{MDS}_{(\alpha+\beta){{N}\choose{K}}\times\alpha{{N}\choose{K}}}$.

$\bullet$ Step 2: Let $L=(\alpha+\beta){N\choose K}$. Independently choose $M$ random matrices, $S_1,\dots,S_M$, uniformly from all the $L\times L$ full rank matrices over $\mathbb{F}_q$.

$\bullet$ Step 3: Build an {\it assisting array} of size ${{N-1}\choose{K-1}}\times N$ consisting of ${{N}\choose{K}}$ symbols. Each symbol appears $K$ times and every $K$ columns share a common symbol.

$\bullet$ Step 4: \emph{[Building basic blocks of the query structure]} The query structure is divided into $M\alpha+P\beta$ blocks.

The $M\alpha$ blocks are denoted as $\Lambda^{[m]}_{\lambda}$, $1\le \lambda \le \alpha$ and $1\le m \le M$. In such a block any query is an atom towards the corresponding file $W^{[m]}$. The atoms appeared in a block $\Lambda^{[m]}_{\lambda}$ are the atoms $a^{[m]}_{[(\beta+\lambda-1){N\choose K}+1:(\beta+\lambda){N\choose K}]}$. The block is set in an ``isomorphic" form with the assisting array, i.e., every $K$ servers share a common atom.

Pick a random MDS generator matrix $\mathbf{H}\in\mathbb{F}_q^{P\times M}$. For example, this could be done by choosing a Reed-Solomon generator matrix and then randomly permuting the columns, resulting in a matrix $\mathbf{H}$ of the form
\begin{equation}
  \mathbf{H}=\left(
               \begin{array}{c}
                 \mathbf{h}_1 \\
                 \mathbf{h}_2 \\
                 \vdots \\
                 \mathbf{h}_P \\
               \end{array}
             \right)=\left(
                       \begin{array}{cccc}
                         h_{11} & h_{12} & \cdots & h_{1M} \\
                         h_{21} & h_{22} & \cdots & h_{2M} \\
                         \vdots & \vdots & \ddots & \vdots \\
                         h_{P1} & h_{P2} & \cdots & h_{PM} \\
                       \end{array}
                     \right).
\end{equation}

The $P\beta$ blocks are denoted as $\Lambda^{\Sigma}_{\lambda,p}$, $1\le p\le P$, $1\le \lambda \le \beta$. In such a block any query is a mixture of $M$ atoms. Again the block is set in an ``isomorphic" form with the assisting array. If the corresponding entry in the assisting array is the number $x\in\{1,2,\dots,{N\choose K}\}$, then we set the corresponding query in the block $\Lambda^{\Sigma}_{\lambda,p}$ to be the formal sum
\begin{equation}
\mathbf{h}_p\cdot (a^{[1]}_{(\lambda-1){N\choose K}+x},a^{[2]}_{(\lambda-1){N\choose K}+x},\dots,a^{[M]}_{(\lambda-1){N\choose K}+x})=\sum_{m=1}^{M}h_{pm}a^{[m]}_{(\lambda-1){N\choose K}+x}.
\end{equation}

$\bullet$ Step 5: \emph{[Constructing the atoms for the desired files]} The atoms for each desired file $W^{[m]}$, $1\le m \le P$ are just built by $a^{[m]}_{[1:L]}=S_mW^{[m]}$.

$\bullet$ Step 6: \emph{[Constructing the atoms for the undesired files]} For each undesired file $W^{[m]}$, choose the first $\alpha{N\choose K}$ rows of $S_m$, denoted as $S_m[(1:\alpha{N\choose K}),:]$. The atoms $a^{[m]}_{[1:L]}$ are built by
\begin{equation}
a^{[m]}_{[1:L]}=\text{MDS}_{(\alpha+\beta){{N}\choose{K}}\times\alpha{{N}\choose{K}}}S_m[(1:\alpha{N\choose K}),:]W^{[m]}.
\end{equation}

The construction of the multi-file PIR scheme for $P\ge\frac{M}{2}$ is finished. Comparing the scheme with the prototype, one can see that the main adaptation is a different query structure. Instead of having blocks labelled by any nonempty subset $\mathcal{D}\subseteq\{W^{[1]},W^{[2]},\dots,W^{[M]}\}$, now we only have blocks labelled by subsets of cardinality 1 and $M$.

\subsubsection{An example: $N=4,K=2,T=2,M=3,P=2$}

Denote the three files by $W^{[1]}$, $W^{[2]}$ and $W^{[3]}$. The parameters throughout the construction can be calculated as $\alpha=5$, $\beta=1$ and $L=36$. Assume each file is of length $72$ and represented in a matrix of size $36\times2$, i.e.,
\begin{equation}
  W^{[1]}=\Bigg(
                  \begin{array}{c}
                    \mathbf{w}^{[1]}_1 \\
                    \vdots \\
                    \mathbf{w}^{[1]}_{36} \\
                  \end{array}
                \Bigg)
  \hspace{3em}
  W^{[2]}=\Bigg(
                  \begin{array}{c}
                    \mathbf{w}^{[2]}_1 \\
                    \vdots \\
                    \mathbf{w}^{[2]}_{36} \\
                  \end{array}
                \Bigg)
                \hspace{3em}
  W^{[3]}=\Bigg(
                  \begin{array}{c}
                    \mathbf{w}^{[3]}_1 \\
                    \vdots \\
                    \mathbf{w}^{[3]}_{36} \\
                  \end{array}
                \Bigg)
\end{equation}
where $\mathbf{w}^{[1]}_i,\mathbf{w}^{[2]}_i,\mathbf{w}^{[3]}_i\in\mathbb{F}_q^{2}$, $1\le i \le 36$. Here $\mathbb{F}_q$ is a sufficiently large finite field. Let $W^{[1]}$ and $W^{[2]}$ be the desired files.

Independently choose three random matrices $S_1,S_2,S_3\in\mathbb{F}_q^{36\times36}$ uniformly from all the $36\times36$ full rank matrices over $\mathbb{F}_q$. Construct a list of {\it atoms} $a^{[1]}_{[1:36]}=S_1W^{[1]}$.  Similarly let $a^{[2]}_{[1:36]}=S_2W^{[2]}$. Suppose there exists a $(36,30)$-MDS code and the transpose of its generator matrix is denoted as $\text{MDS}_{36\times30}$. Select the first 30 rows of $S_3$, denoted as $S_3[(1:30),:]$. Construct a list of atoms $a^{[3]}_{[1:36]}=\text{MDS}_{36\times30}S_3[(1:30),:]W^{[3]}$.

The atoms $a^{[1]}_{[1:36]}$, $a^{[2]}_{[1:36]}$ and $a^{[3]}_{[1:36]}$ will form the queries for the servers. The queries for the servers are divided into seventeen blocks. In fifteen blocks each query is only a single atom and in two blocks $\Lambda^{\Sigma}_1$ and $\Lambda^{\Sigma}_2$ each query is a mixture of three atoms. The query structure is as follows.

\medskip

\noindent\fbox{%
  \parbox{\textwidth}{%
$$\Lambda^{\Sigma}_1:\Bigg\{\begin{array}{cccc}
  \text{Server I} & \text{Server II} & \text{Server III} & \text{Server IV}\\\hline
  a^{[1]}_{1}+a^{[2]}_{1}+a^{[3]}_{1} & a^{[1]}_{1}+a^{[2]}_{1}+a^{[3]}_{1} & a^{[1]}_{2}+a^{[2]}_{2}+a^{[3]}_{2} & a^{[1]}_{2}+a^{[2]}_{2}+a^{[3]}_{2} \\
  a^{[1]}_{3}+a^{[2]}_{3}+a^{[3]}_{3} & a^{[1]}_{4}+a^{[2]}_{4}+a^{[3]}_{4} & a^{[1]}_{3}+a^{[2]}_{3}+a^{[3]}_{3} & a^{[1]}_{4}+a^{[2]}_{4}+a^{[3]}_{4} \\
  a^{[1]}_{5}+a^{[2]}_{5}+a^{[3]}_{5} & a^{[1]}_{6}+a^{[2]}_{6}+a^{[3]}_{6} & a^{[1]}_{6}+a^{[2]}_{6}+a^{[3]}_{6} & a^{[1]}_{5}+a^{[2]}_{5}+a^{[3]}_{5} \\
\end{array}\Bigg\}.$$

$$\Lambda^{\Sigma}_2:\Bigg\{\begin{array}{cccc}
  \text{Server I} & \text{Server II} & \text{Server III} & \text{Server IV}\\\hline
  z_1 a^{[1]}_{1}+z_2 a^{[2]}_{1}+z_3 a^{[3]}_{1} & z_1 a^{[1]}_{1}+z_2 a^{[2]}_{1}+z_3 a^{[3]}_{1} & z_1 a^{[1]}_{2}+z_2 a^{[2]}_{2}+z_3 a^{[3]}_{2} & z_1 a^{[1]}_{2}+z_2 a^{[2]}_{2}+z_3 a^{[3]}_{2} \\
  z_1 a^{[1]}_{3}+z_2 a^{[2]}_{3}+z_3 a^{[3]}_{3} & z_1 a^{[1]}_{4}+z_2 a^{[2]}_{4}+z_3 a^{[3]}_{4} & z_1 a^{[1]}_{3}+z_2 a^{[2]}_{3}+z_3 a^{[3]}_{3} & z_1 a^{[1]}_{4}+z_2 a^{[2]}_{4}+z_3 a^{[3]}_{4} \\
  z_1 a^{[1]}_{5}+z_2 a^{[2]}_{5}+z_3 a^{[3]}_{5} & z_1 a^{[1]}_{6}+z_2 a^{[2]}_{6}+z_3 a^{[3]}_{6} & z_1 a^{[1]}_{6}+z_2 a^{[2]}_{6}+z_3 a^{[3]}_{6} & z_1 a^{[1]}_{5}+z_2 a^{[2]}_{5}+z_3 a^{[3]}_{5} \\
\end{array}\Bigg\}$$
where $z_1,z_2,z_3$ are randomly chosen distinct elements in $\mathbb{F}_q$.

$$\Lambda^{[x]}_{\lambda}:\Bigg\{\begin{array}{cccc}
  \text{Server I} & \text{Server II} & \text{Server III} & \text{Server IV}\\\hline
  a^{[x]}_{6\lambda+1} & a^{[x]}_{6\lambda+1} & a^{[x]}_{6\lambda+2} & a^{[x]}_{6\lambda+2} \\
  a^{[x]}_{6\lambda+3} & a^{[x]}_{6\lambda+4} & a^{[x]}_{6\lambda+3} & a^{[x]}_{6\lambda+4} \\
  a^{[x]}_{6\lambda+5} & a^{[x]}_{6\lambda+6} & a^{[x]}_{6\lambda+6} & a^{[x]}_{6\lambda+5} \\
\end{array}\Bigg\}$$
for $\lambda\in\{1,2,3,4,5\}$, $x\in\{1,2,3\}$.
  }%
}

\medskip

Retrieving the files $W^{[1]}$ and $W^{[2]}$ is equivalent to retrieving $a^{[1]}_{[1:36]}$ and $a^{[2]}_{[1:36]}$ since $S_1$ and $S_2$ are both full rank matrices. $a^{[1]}_{[7:36]}$, $a^{[2]}_{[7:36]}$ and $a^{[3]}_{[7:36]}$  could be retrieved. Then $a^{[3]}_{[1:6]}$ could be solved from $a^{[3]}_{[7:36]}$ due to the property of $\text{MDS}_{36\times30}$. Therefore, the interferences can be eliminated in the blocks $\Lambda^{\Sigma}_1$ and $\Lambda^{\Sigma}_2$. So the queries of the form $a^{[1]}_i+a^{[2]}_i$ and $z_1 a^{[1]}_i+z_2 a^{[2]}_i$, $1\le i \le6$, could be retrieved. Then we can solve the six systems of equations to retrieve $a^{[1]}_{[1:6]}$ and $a^{[2]}_{[1:6]}$.

The scheme is private against any two colluding servers. From the perspective of any two colluding servers, altogether they have 30 distinct atoms towards each file. Extract the coefficients of each atom as a vector in $\mathbb{F}_q^{36}$. Recall how we select the random matrices $S_1$, $S_2$ $S_3$ and the $(36,30)$-MDS code. It turns out that the 30 vectors with respect to each file form a random subspace of dimension 30 in $\mathbb{F}_q^{36}$, so the two servers cannot tell any difference among the atoms towards different files and thus the identity of the retrieved files is disguised.

The rate of the scheme above is then $\frac{36\times2\times2}{12\times5\times3+12+12}=\frac{12}{17}$. As a comparison, for the sake of retrieving two files a trivial approach is to execute two independent (single-file) PIR schemes as shown in the example in Section \ref{Prototype} (note that 25 extra bits of $W^{[2]}$ could be retrieved in the scheme designed for retrieving 36 bits of $W^{[1]}$ out of the 91 downloaded bits) and the rate is only $\frac{61}{91}<\frac{12}{17}$.

\subsubsection{Analysis of the multi-file PIR scheme for $P\ge\frac{M}{2}$}

Each query appears in $K$ different servers and thus can be successfully retrieved. So we have retrieved all the atoms $a^{[m]}_{[\beta{N\choose K}+1:L]}$ for each file $W^{[m]}$. Then by the property of $\text{MDS}_{(\alpha+\beta){{N}\choose{K}}\times\alpha{{N}\choose{K}}}$, we can solve $a^{[m]}_{[1:\beta{N\choose K}]}$ from $a^{[m]}_{[\beta{N\choose K}+1:L]}$ for each undesired file. After eliminating these interferences from the queries in those $P\beta$ blocks $\Lambda^{\Sigma}_{\lambda,p}$, we are facing with $\beta$ systems of equations. The coefficients of each system come from some $P$ columns of $\mathbf{H}$ and therefore form a non-singular $P\times P$ matrix. By solving these systems we retrieve $a^{[m]}_{[1:\beta{N\choose K}]}$ for every desired file. Therefore for every desired file we have retrieved all the $L$ atoms towards the file, which is equivalent to the retrieval of the file itself since each $S_m$ is a full rank matrix.

The scheme is private against any $T$ colluding servers. From the perspective of any $T$ colluding servers, the query structure is completely symmetric with respect to all the files. The number of atoms towards any file known by these $T$ colluding servers is $(\alpha+\beta)\big({{N}\choose{K}}-{{N-T}\choose{K}}\big)$. Extract the coefficients of an atom as a vector in $\mathbb{F}_q^{L}$. Recall how we build the atoms and the property of the $\big((\alpha+\beta){{N}\choose{K}},\alpha{{N}\choose{K}}\big)$-MDS code. One can see that the vectors with respect to each file will form a random subspace of dimension $\alpha{{N}\choose{K}}$ in $\mathbb{F}_q^{L}$. So these $T$ colluding servers cannot tell any difference among the atoms towards different files and thus the identity of the retrieved files is disguised.

The rate of the scheme can be calculated as
\begin{equation}
\frac{PLK}{(M\alpha+P\beta)K{N\choose K}}=\frac{P\alpha+P\beta}{M\alpha+P\beta}=\frac{{N\choose K}}{\frac{M}{P}\big({N\choose K}-{{N-T}\choose K}\big)+{{N-T}\choose K}}.
\end{equation}

Recall that we are now dealing with the case $P\ge\frac{M}{2}$. When $T=1$, the rate above is exactly $\big(1+\frac{K(M-P)}{PN}\big)^{-1}$, meeting the upper bound of the capacity shown in (\ref{CapacityK2}). When $K=1$, the rate above is exactly $\big(1+\frac{T(M-P)}{PN}\big)^{-1}$, meeting the upper bound of the capacity shown in (\ref{CapacityT2}). Therefore, when $P\ge\frac{M}{2}$, we determine the exact multi-file PIR capacity for the degenerating cases. Again the condition $T+K\le N$ is essential since otherwise the equality (\ref{equationmulti1}) does not make sense. We summarize our main result as follows.

\begin{theorem}
  There exists an $(N,K,T;M)$ $P$-files PIR scheme with rate $\frac{{N\choose K}}{\frac{M}{P}\big({N\choose K}-{{N-T}\choose K}\big)+{{N-T}\choose K}}$ when $T+K\le N$ and $P\ge\frac{M}{2}$. The rate of the scheme is optimal when either $K=1$ or $T=1$.
\end{theorem}

We close this subsection by briefly discussing the naive approach of executing $P$ independent (single-file) PIR schemes. Using our general (single-file) scheme, we can only have a $P$-file PIR scheme with rate
\begin{equation}
(1+R^{M-1}P-R^{M-1})(1+R+R^2+\cdots+R^{M-1})^{-1},
\end{equation}
where $R=1-\frac{{{N-T}\choose K}}{{N\choose K}}$. Within this calculation one should take into consideration that we can retrieve a certain amount of each of the other $P-1$ files for free when we execute a (single-file) PIR scheme towards a fixed file. To check that the adapted multi-file PIR scheme performs better than the trivial approach is equivalent to showing
\begin{equation}
1+R+\cdots+R^{M-1}>(\frac{M}{P}R+1-R)(1+R^{M-1}P-R^{M-1}).
\end{equation}
The right-hand side achieves maximum at either $P=M$ (then its value is $1+(M-1)R^{M-1}$) or $P=\frac{M}{2}$ (then its value is $(1+R)(1+(\frac{M}{2}-1)R^{M-1})$), both smaller than the left-hand side. So the adapted multi-file PIR scheme is indeed better than the trivial approach.

\section{PIR with arbitrary collusion patterns} \label{SecArbitrary}

In this section we deal with the last variant model, PIR with arbitrary collusion patterns. In this model the files are stored via an arbitrary $(N,K)$-MDS code as usual. We have a collusion pattern $\mathfrak{T}$, which is a family of subsets of the set of servers. A {\it collusion subset} $\mathcal{T}\in\mathfrak{T}$ indicates that this set of servers may collude in attacking on the identity of the desired file. Naturally, any subset of a collusion set is again a collusion set, and therefore $\mathfrak{T}$ is closed under inclusion. Thus {\it collusion pattern} $\mathfrak{T}$ could be denoted by its maximal collusion sets, say $\mathfrak{T}=<\mathcal{T}_1,\mathcal{T}_2,\dots,\mathcal{T}_r>$. Designing PIR schemes in this model may follow a case-by-case analysis due to the property of the specific collusion pattern. Suppose the cardinality of the largest collusion set is $T$, then a PIR scheme against any $T$ colluding servers certainly works. However, there is some space for improvements.

To adapt our general PIR scheme for this model, the trick is to adjust the assisting array used in the construction. First we need to analyze the following optimization problem, which is closely related to combinatorial design theory. A family of $K$-subsets of the servers is denoted as $\mathfrak{B}=\{\mathcal{B}_1,\mathcal{B}_2,\dots,\mathcal{B}_b\}$. For every maximal collusion set $\mathcal{T}_i$, $1\le i \le r$, let $\#(\mathcal{T}_i)$ denote the number of $K$-subsets in $\mathfrak{B}$ which have non-empty intersections with $\mathcal{T}_i$. Let $\Delta=\max\{\#(\mathcal{T}_i):1\le i \le r\}$. The problem is to find a proper family $\mathfrak{B}$ to minimize $\Delta/b$. For example, when we consider the previous models where $\mathfrak{T}$ denotes every $T$-subset of the servers, we simply set $\mathfrak{B}$ to be the collection of every $K$-subset of the servers. Then $b={N\choose K}$ and $\Delta={N\choose K}-{{N-T}\choose K}$.

\subsection{PIR scheme with arbitrary collusion patterns}

Constructing the PIR scheme with arbitrary collusion patterns contains the following steps.

$\bullet$ Step 0: Given the collusion pattern $\mathfrak{T}$, find a proper family $\mathfrak{B}$ to solve the optimization problem above.

$\bullet$ Step 1: Let $\alpha$ and $\beta$ be the smallest positive integers satisfying
\begin{equation} \label{equation}
\alpha b=(\alpha+\beta)\Delta.
\end{equation}
Assume the existence of an $\big((\alpha+\beta)b,\alpha b\big)$-MDS code. The transpose of its generator matrix is $\text{MDS}_{(\alpha+\beta)b\times\alpha b}$.

$\bullet$ Step 2: Let $L=b(\alpha+\beta)^{M-1}$.  Independently choose $M$ random matrices, $S_1,\dots,S_M$, uniformly from all the $L\times L$ full rank matrices over $\mathbb{F}_q$. 

$\bullet$ Step 3: Build an {\it assisting array} according to the choice of $\mathfrak{B}$. Each symbol appears $K$ times and every subset of servers $\mathcal{B}\in\mathfrak{B}$ share a common symbol.

$\bullet$ Step 4: \emph{[Construction of the query structure]} The query structure is divided into several blocks, where each block is labelled by $\mathcal{D}\subseteq\{W^{[1]},W^{[2]},\dots,W^{[M]}\}$, a subset of files. In a block labelled by $\mathcal{D}$, the queries are of the same form, i.e., every query is a mixture of $|\mathcal{D}|$ atoms related to the files in $\mathcal{D}$. Set each block in an ``isomorphic" form with the assisting array. We further call a block labelled by $\mathcal{D}$ a {\it $d$-block} if $|\mathcal{D}|=d$. For any $\mathcal{D}\subseteq\{W^{[1]},W^{[2]},\dots,W^{[M]}\}$, $|\mathcal{D}|=d$, we require that the number of $d$-blocks labelled by $\mathcal{D}$ is $\alpha^{M-d}\beta^{d-1}$.

$\bullet$ Step 5: \emph{[Constructing the atoms for the desired file in each block]} The atoms for the desired file $W^{[1]}$ are just built by $S_1W^{[1]}$ and then distributed to different blocks.

$\bullet$ Step 6: \emph{[Dividing the query structure into groups]} Now let $\mathcal{D}$ denote a nonempty set of files not containing the desired file $W^{[1]}$. There are totally $\alpha^{M-|\mathcal{D}|}\beta^{|\mathcal{D}|-1}$ blocks labelled by $\mathcal{D}$ and $\alpha^{M-|\mathcal{D}|-1}\beta^{|\mathcal{D}|}$ blocks labelled by $\mathcal{D}\bigcup\{W^{[1]}\}$. Partition these blocks into $\alpha^{M-|\mathcal{D}|-1}\beta^{|\mathcal{D}|-1}$ groups, denoted by $\Gamma^{\mathcal{D}}_{\lambda}$, $1\le\lambda\le \alpha^{M-|\mathcal{D}|-1}\beta^{|\mathcal{D}|-1}$, where each group consists of $\alpha$ blocks labelled by $\mathcal{D}$ and $\beta$ blocks labelled by $\mathcal{D}\bigcup\{W^{[1]}\}$. Repeat this process for every nonempty set $\mathcal{D}$ not containing $W^{[1]}$ and thus the whole query structure is partitioned into groups.

$\bullet$ Step 7: \emph{[Constructing the atoms for undesired files in each group]} For each group constructed with respect to $\mathcal{D}$ and any file $W^{[m]}\in\mathcal{D}$, $m\neq1$, the number of atoms towards $W^{[m]}$ within the blocks in this group is $(\alpha+\beta)b$, among which $\alpha b$ atoms appear in blocks labelled by $\mathcal{D}$ and the other $\beta b$ atoms appear as interferences in blocks labelled by $\mathcal{D}\bigcup\{W^{[1]}\}$. Let these $(\alpha+\beta)b$ atoms for $W^{[m]}$ be built by $\text{MDS}_{(\alpha+\beta)b\times\alpha b}S'_mW^{[m]}$, where $S'_m$ denotes some $\alpha b$ rows in the matrix $S_m$.

Repeat this process for all the groups. Note that we shall come across the same file $W^{[M]}$ several times. We have ${{M-2}\choose{|\mathcal{D}|-1}}$ choices for a subset $\mathcal{D}$ of size $|\mathcal{D}|$, containing $W^{[M]}$ but without $W^{[1]}$. With respect to any such $\mathcal{D}$ we have $\alpha^{M-|\mathcal{D}|-1}\beta^{|\mathcal{D}|-1}$ groups. So the exact number of times we come across $W^{[M]}$ can be computed as
\begin{equation}
  \sum_{|\mathcal{D}|=1}^{M-1} \alpha^{M-|\mathcal{D}|-1}\beta^{|\mathcal{D}|-1}{{M-2}\choose{|\mathcal{D}|-1}}=(\alpha+\beta)^{M-2}.
\end{equation}

Every time we come across the same file $W^{[M]}$, we have to ensure that the rows we select from $S_m$ are non-intersecting. This can be satisfied since $(\alpha+\beta)^{M-2}\alpha b\le L$.

The construction of the PIR scheme with respect to the particular collusion pattern $\mathfrak{T}$ is finished. Comparing the scheme with the prototype, one can see that the main adaptation is changing the assisting array and then adjusting the ratio of different blocks according to the new assisting array.

\subsection{An example}

Let $N=5$, $K=3$, $M=2$. The collusion pattern is $\mathfrak{T}=\{\{I,II\},\{II,III\},\{III,IV\},\{IV,V\},\{I,V\}\}$. Instead of choosing the collection of every $3$-subset, we could set $\mathfrak{B}=\{\{I,II,III\},\{II,III,IV\},\{III,IV,V\},\{I,IV,V\},\{I,II,V\}\}$ and build the corresponding assisting array such that every subset of servers $B\in\mathfrak{B}$ share a common entry:
$$\Bigg\{\begin{array}{ccccc}
  1 & 2 & 3 & 4 & 5\\
  5 & 1 & 2 & 3 & 4\\
  4 & 5 & 1 & 2 & 3\\
\end{array}\Bigg\}.$$

The parameters throughout the construction can be calculated as $b=5$, $\Delta=4$, $\alpha=4$, $\beta=1$ and then $L=25$.

Now we have two files $U$ and $V$. Let each file be of length $75$ and represented in a matrix of size $25\times3$, i.e.,
\begin{equation}
  U=\Bigg(
                  \begin{array}{c}
                    \mathbf{u}_1 \\
                    \vdots \\
                    \mathbf{u}_{25} \\
                  \end{array}
                \Bigg)
  \hspace{3em}
  V=\Bigg(
                  \begin{array}{c}
                    \mathbf{v}_1 \\
                    \vdots \\
                    \mathbf{v}_{25} \\
                  \end{array}
                \Bigg)
\end{equation}
where $\mathbf{u}_i,\mathbf{v}_i\in\mathbb{F}_q^{3}$, $1\le i \le 25$. Here $\mathbb{F}_q$ is a sufficiently large finite field. Let $U$ be the desired file. We shall construct two lists of atoms $a_{[1:25]}$ and $b_{[1:25]}$ corresponding to each file accordingly. Following Step 4, the query structure is divided into several blocks as follows. In each block we set the queries in an ``isomorphic" form with the assisting array.

$$\Lambda^{UV}:\Bigg\{\begin{array}{ccccc}
  \text{Server I} & \text{Server II} & \text{Server III} & \text{Server IV} & \text{Server V} \\\hline
  a_{1}+b_{1} & a_{2}+b_{2} & a_{3}+b_{3} & a_{4}+b_{4} & a_{5}+b_{5} \\
  a_{5}+b_{5} & a_{1}+b_{1} & a_{2}+b_{2} & a_{3}+b_{3} & a_{4}+b_{4} \\
  a_{4}+b_{4} & a_{5}+b_{5} & a_{1}+b_{1} & a_{2}+b_{2} & a_{3}+b_{3} \\
\end{array}\Bigg\}\text{,}$$

$$\Lambda^X_{\eta}:\Bigg\{\begin{array}{ccccc}
  \text{Server I} & \text{Server II} & \text{Server III} & \text{Server IV} & \text{Server V} \\\hline
x_{5\eta+1} & x_{5\eta+2} & x_{5\eta+3} & x_{5\eta+4} & x_{5\eta+5} \\
x_{5\eta+5} & x_{5\eta+1} & x_{5\eta+2} & x_{5\eta+3} & x_{5\eta+4} \\
x_{5\eta+4} & x_{5\eta+5} & x_{5\eta+1} & x_{5\eta+2} & x_{5\eta+3} \\
\end{array}\Bigg\}\text{ for $(x,X)\in\{(a,U),(b,V)\}$ and $\eta\in\{1,\cdots,4\}$.}$$

The remaining task is then to determine the atoms. In Step 2 we have a $(25,20)$-MDS code and the transpose of its generator matrix is denoted as $\text{MDS}_{25\times20}$.  Independently choose two random matrices $S_1,S_2\in\mathbb{F}_q^{25\times25}$, uniformly from all the $25\times25$ full rank matrices over $\mathbb{F}_q$. The atoms $a_{[1:25]}$
are just built by $S_1U$. The atoms $b_{[1:25]}$ are built by $\text{MDS}_{25\times20}S_2[(1:20),:]V$.

Retrieving the file $U$ is equivalent to retrieving $a_{[1:25]}$ since $S_1$ is a full rank matrix. Retrieving $a_{[6:25]}$ and $b_{[6:25]}$ is straightforward. Using the MDS property of $\text{MDS}_{25\times20}$, $b_{[1:5]}$ can be solved from $b_{[6:25]}$. Eliminating these interferences we now retrieve $a_{[1:5]}$.

The scheme is private against the specific collusion pattern. The query structure is completely symmetric with respect to all the files. From the perspective of any two possible colluding servers, altogether the number of atoms towards each of the three files is $20$. Extract the coefficients of each atom as a vector in $\mathbb{F}_q^{25}$. Recall how we select the random matrices $S_1$, $S_2$, and the $(25,20)$-MDS code. It turns out that the 20 vectors with respect to each file form a random subspace of dimension 20 in $\mathbb{F}_q^{25}$, so the two servers cannot tell any difference among the atoms towards different files and thus the identity of the retrieved file is disguised.

Note that if Server I and Server III collude, then the scheme fails. Server I and Server III know all the atoms. They will discover that the 25 atoms towards $U$ are independent and the 25 atoms towards $V$ have certain redundancy. Therefore they will realize that $U$ is the desired file.

The rate of the scheme above is then $\frac{25\times3}{3\times5\times9}=\frac{5}{9}$. This improves on the result $\frac{10}{19}$ if we use a general scheme against any two colluding servers.

\subsection{Analysis of the PIR scheme with arbitrary collusion patterns}

Each query in the scheme appears in $K$ distinct servers and is thus retrieved. For each group constructed with respect to $\mathcal{D}$, using the property of $\text{MDS}_{(\alpha+\beta)b\times\alpha b}$, we can use the $\alpha b$ queries in the blocks labelled by $\mathcal{D}$ to eliminate the interferences in the $\beta b$ blocks labelled by $\mathcal{D}\bigcup\{W^{[1]}\}$. Therefore all the atoms towards $W^{[1]}$ can be retrieved. By the singularity of $S_1$, the desired file is successfully retrieved.

The scheme is private against any set of possible colluding servers. For a maximal colluding set $\mathcal{T}\in\mathfrak{T}$, the atoms towards any file known by these servers is $(\alpha+\beta)^{M-1}\Delta$. Extract the coefficients of each atom as a vector in $\mathbb{F}_q^{L}$. Recall how we build the atoms by using the full rank matrices $S_1,\dots,S_m$ and the MDS code. One can see that the vectors of all the atoms towards any file will form a random subspace of dimension $(\alpha+\beta)^{M-1}\Delta$ in $\mathbb{F}_q^{L}$. So these colluding servers cannot tell any difference among the atoms towards different files and thus the identity of the retrieved file is disguised.

We compute the rate of the PIR scheme. The retrieved file is of length $LK$. The download cost in any block is $bN$. So the rate can be computed as
$$\frac{LK}{bK\sum_{|\mathcal{F}|=1}^{M} \alpha^{M-|\mathcal{F}|}\beta^{|\mathcal{F}|-1}{{M}\choose{|\mathcal{F}|}}}=\frac{L}{b\beta^{-1}\big((\alpha+\beta)^M-\alpha^M\big)}=\frac{1}{1+\frac{\alpha}{\alpha+\beta}+\cdots+(\frac{\alpha}{\alpha+\beta})^{M-1}}.$$

Similarly as before, a final remark is that our scheme only works for the case $b>\Delta$ since otherwise the equality (\ref{equationByzantine}) does not make sense (This can be satisfied as long as $T+K\le N$, where $T$ is the cardinality of a maximal colluding set). To sum up, we have the following theorem.

\begin{theorem}
  Given an arbitrary collusion pattern $\mathfrak{T}$, suppose we can build a family of $K$-subsets of servers $\mathfrak{B}=\{\mathcal{B}_1,\mathcal{B}_2,\dots,\mathcal{B}_b\}$ such that any $\mathcal{T}\in\mathfrak{T}$ intersects with at most $\Delta$ subsets in $\mathfrak{B}$. If $b>\Delta$, then there exists an $(N,K;M)$-PIR scheme with respect to this collusion pattern whose rate is $(1+R+R^2+\cdots+R^{M-1})^{-1}$, where $R=\frac{\Delta}{b}$.
\end{theorem}

\section{Conclusion} \label{SecConclusion}

In this paper, we adapt our general PIR scheme presented in \cite{Zhangnew} for four variant PIR models. The idea can be generalized (although maybe tedious) further to some hybrid models, say a multi-file PIR model with both robust and Byzantine servers. The results contain a series of previous works as special cases. However, it is not clear to evaluate the performance of the scheme for non-degenerating parameters $K$ and $T$, since the exact PIR capacity (or even only a good upper bound) remains unknown. A first step to analyze the PIR capacity for the general model will be of great interest.

\end{document}